\documentclass[numberedappendix]{emulateapj}
\usepackage{apjfonts,mathptmx}

\begin{document}

\title{Spectral Methods for Time-Dependent Studies of Accretion Flows.\\
       I. Two-Dimensional, Viscous, Hydrodynamic Disks}

\author{Chi-kwan Chan, Dimitrios Psaltis, and Feryal \"Ozel\altaffilmark{1}}
\affil{Physics and Astronomy Departments, University of Arizona, 1118 E.\ 4th St., Tucson, AZ 85721}

\altaffiltext{1}{Hubble Fellow}

\begin{abstract}
We present a numerical method for studying the normal modes of accretion flows around black holes. In this first paper,
we focus on two-dimensional, viscous, hydrodynamic disks, for which the linear modes have been calculated analytically
in previous investigations. We use pseudo-spectral methods and low storage Runge-Kutta methods to solve the continuity
equation, the Navier-Stokes equation, and the energy equation. We devise a number of test problems to verify the
implementation. These tests demonstrate the ability of spectral methods to handle accurately advection problems and to
reproduce correctly the stability criteria for differentially rotating hydrodynamic flows. They also show that our
implementation is able to handle sound wave correctly with non-reflective boundary conditions, to recover the standard
solution for a viscous spreading ring, and produce correctly the Shakura-Sunyaev steady disk solution. Finally, we have
applied our algorithm to the problem of a non-axisymmetric viscous spreading ring and verify that such configuration is
unstable to non-antisymmetric perturbations.
\end{abstract}

%---------------------------------------------------------------------------------------------------------------------

\section{Introduction}

Observations of accretion flows around black holes in galactic systems and in the centers of galaxies show strong
evidence for the presence of long-lived, global modes at characteristic frequencies of the black-hole spacetimes
\citep[see][for reviews]{VanDerKlis2000,McClintock2004}. The origin of these modes is still a matter of debate
\citep[see, e.g.,][]{Psaltis2001,Psaltis2004}. It is anticipated, however, that modeling accurately their origin and
physical properties will provide the first measurements of the spins of black holes in the universe as well as direct
tests of strong-field general relativity \citep{Psaltis2004}.

Systematic studies of modes in accretion disks around black holes have so far been mostly analytical, performed in the
linear regime for flows with an artificial viscosity law \citep[see][and reference therein; see, however, Gammie \&
Balbus 1994 and Menou 2003 for studies of MHD disks]{Wagoner1999,Kato2001}. These studies have shown the presence of
three types of trapped global modes, namely inertial-gravity, corrugation, and inertial-pressure modes, with properties
that appear to agree with observations \citep[see][]{Wagoner2001}.

Despite their success, analytical models have several limitations. First, they rely on a number of approximations, such
as the WKB approximation in the radial direction. Second, because only linear terms in perturbations are taken into
account, these studies are unable to reproduce resonances and couplings between various modes. However, the detected
QPOs often have fractional amplitudes as large as 40\% of the total X-ray flux, making the linear approximation
inaccurate \citep[see, e.g.,][]{McClintock2004}. Furthermore, pairs of QPOs are typically observed
\citep[see][]{Strohmayer2001a,Strohmayer2001b} with frequency ratios equal to ratios of small integers, strongly
arguing for the presence of resonances between modes \citep{Abramowicz2001}. Third, analytical and numerical studies of
oscillations in MHD disks cast doubt on the presence of oscillations at the radial epicyclic frequencies, as required
by all diskoseismic calculations \citep[see, e.g.,][]{Hawley2001,Hawley2002,Armitage2003}. Answering all the above
questions necessitates a systematic, numerical study of accretion disk oscillations.

Numerical studies of accretion disk modes have been very few and have not addressed in detail the above issues
\citep[see, e.g.,][for simulations of viscous disks; see also Rezzolla 2004 for simulations of hydrodynamic
tori]{Chen1994,Chen1995,Milsom1996,Milsom1997}. In fact, no numerical simulation of hydrodynamic disks to date has
resulted in solutions that show all three types of accretion-disk modes that were discovered analytically. In this
series of papers, we perform a systematic study of the oscillatory behavior of hydrodynamic and MHD accretion disks in
two and three dimensions. To this end, we have developed a pseudo-spectral numerical algorithm, which we present here.
Even though spectral methods have been used extensively in the study of hydrodynamic modes in various planetary
atmospheres, including the earth's \citep[see, e.g.,][]{Mote2000}, their application to the studies of accretion disks
has been very limited \citep{Godon1997}.

Spectral algorithms are high order numerical methods and are perfectly suited to studying modes of hydrodynamic flows
for a number of reasons. Their major advantage is accuracy and economy in the number of degrees of freedom
\citep{Bonazzola1999}. In spectral methods, only $\sim\pi$ collocation points are needed to resolve one wavelength
\citep{Boyd2000}, whereas in finite difference methods a lot more grid points are required to reach the same accuracy.
The solution is already expressed as a set of orthogonal modes; this property improves the accuracy of studies of mode
coupling and resonance. Moreover, the dimensionality of the problem can be altered very easily, for investigating the
effects of the number of spatial dimensions on the properties of hydrodynamic modes and turbulence. In simulating MHD
flows, the high order of spectral methods allows for the equations to be solved in terms of the vector potential and
not of the magnetic field itself; this guarantees that the resulting magnetic field is always divergence free. Also,
super spectral viscosity \citep{Ma1998a,Ma1998b}, which is one kind of artificial viscosity, is trivial to implement in
spectral methods. It resolves shocks and yet preserves the high spectral accuracy of the solution. Finally,
incorporating the self-gravity of the flows is trivial and does not increase significantly the computational time
\citep[see, e.g.,][]{Chen2000}.

In this first paper, we discuss our numerical algorithm for evolving two-dimensional viscous, hydrodynamic accretion
disks. In the following section, we present our assumptions and equations. In \S3, we discuss the details of our
numerical method. In \S4, we present a series of tests to verify our algorithm. In \S5, we apply our algorithm to
non-axisymmetric instabilities in viscous spreading rings. We confirm the result of \citet{Speith2003}, which shows
that the standard solution of a viscous spreading ring \citep{Pringle1981,Frank2002} is unstable to non-axisymmetric
perturbation. In \S6, we present our conclusions.

%---------------------------------------------------------------------------------------------------------------------

\section{Equations and Assumptions}\label{sec:equations}

We consider two-dimensional, viscous, compressible flows. In this first paper, we neglect self-gravity and the magnetic
fields of the flows. The hydrodynamic equations, therefore, contain the continuity equation
\begin{equation}
  \frac{\partial\Sigma}{\partial t} + \nabla\cdot(\Sigma\mathbf{v}) = 0, \label{eq:continuity}
\end{equation}
the Navier-Stokes equation
\begin{equation}
  \Sigma\frac{\partial\mathbf{v}}{\partial t} + \Sigma(\mathbf{v}\cdot\nabla)\mathbf{v}
  = -\nabla P + \nabla\mathbf\tau - \Sigma\;\mathbf{g}, \label{eq:navier_stokes}
\end{equation}
and the energy equation
\begin{equation}
  \frac{\partial E}{\partial t} + \nabla\cdot(E\mathbf{v}) = - P\,\nabla\cdot\mathbf{v} + \Phi - \nabla\cdot\mathbf{q} -
  \nabla\cdot\mathbf{F} - 2F_z. \label{eq:energy}
\end{equation}
We denote by $\Sigma$ the height-integrated density, by $\mathbf{v}$ the velocity, and by $E$ the thermal energy. In
the Navier-Stokes equation, $P$ is the height-integrated pressure, $\mathbf\tau$ is the viscosity tensor, and
$\mathbf{g}$ is the gravitational acceleration. We use $\Phi$ to denote the viscous dissipation rate, $\mathbf{q}$ to
denote the heat flux vector, and $\mathbf{F}$ to denote the radiation flux on the $r$-$\phi$ plane. The last term in
the heat equation, $2F_z$, takes into account the radiation losses in the vertical direction.

We allow for large flexibility in the physical assumptions in our algorithm, i.e., we can change the functional form of
$P$, $\mathbf{g}$, $\mathbf{q}$, $\mathbf{F}$, and $F_z$ easily and apply our algorithm to various problems. We also
write the viscosity tensor in the general form
\begin{equation}
  \tau_{i\!j} = 2(\mu_\mathrm{r} + \mu_\mathrm{s})e_{i\!j} + \left(\mu_\mathrm{r} + \mu_\mathrm{b}
  - \frac{2}{3}\mu_\mathrm{s}\right)\nabla\cdot\mathbf{v}.
\end{equation}
Here $\mu_\mathrm{r}$, $\mu_\mathrm{b}$, and $\mu_\mathrm{s}$ are the coefficients of radiative, bulk, and shear
viscosity, respectively. These coefficients can also be changed easily to satisfy different physical assumptions. The
strain-rate tensor $e_{i\!j}$ is
\begin{equation}
  e_{i\!j} = \frac{1}{2}\left(\frac{\partial v_i}{\partial x_j} + \frac{\partial v_j}{\partial x_i}\right)\;.
\end{equation}
The viscous dissipation rate is
\begin{equation}
  \Phi = 2(\mu_\mathrm{r} + \mu_\mathrm{s})(e_{i\!j})^2 + \left(\mu_\mathrm{r} + \mu_\mathrm{b}
  - \frac{2}{3}\mu_\mathrm{s}\right)(\nabla\cdot\mathbf{v})^2\;.
\end{equation}
In appendix~\ref{app:equations}, we write down explicitly the equations used in the algorithm.

The default setup for our calculations is that of a geometrically thin disk, in which radiation pressure is assumed to
be negligible. Using the ideal gas law, the thermal energy density and pressure become
\begin{eqnarray}
  E & = & \Sigma\frac{3k_\mathrm{B}T}{2\mu m_\mathrm{H}}, \\
  P & = & \Sigma\frac{k_\mathrm{B}T}{\mu m_\mathrm{H}},
\end{eqnarray}
where $k_\mathrm{B}$ is the Boltzmann constant, $T$ is the central temperature, $\mu$ is the mean molecular weight, and
$m_\mathrm{H}$ is the mass of the hydrogen atom. The viscosity coefficients $\mu_\mathrm{r}$, $\mu_\mathrm{b}$, and
$\mu_\mathrm{s}$ are chosen based on the physical assumptions of the specific problems.

In order to approximate the effects of general relativity in the vicinity of compact objects, we use the
pseudo-Newtonian approximation for the gravitational acceleration \citep{Paczynsky1980,Mukhopadhyay2002}
\begin{equation}
  \mathbf{g} = \frac{c^2}{r^3}\left[\frac{GM}{c^2}\right]^2\left[\frac{r^2 - 2(a/c)\sqrt{GMr/c^2} + (a/c)^2}
               {\sqrt{GMr/c^2}(r - r_\mathrm{g}) + a/c}\right]^2\mathbf{\hat r}.
\end{equation}
Here, $c$ is speed of light, $G$ is the gravitational constant, $M$ is the mass of the central object, $a$ is its
specific angular momentum, $r_\mathrm{g} = 2GM/c^2$ is its Schwarzschild radius, and $\mathbf{\hat r}$ is the unit
vector in the radial direction. For a non-rotating star, we set $a = 0$ and the gravitational acceleration reduces to
the standard pseudo-Newtonian acceleration $g=GM/(r-r_\mathrm{g})^2$. When $a = r_\mathrm{g} = 0$, it recovers
Newtonian gravity.

Regarding radiative cooling, we follow \citet{Hubeny1990} and \citet{Popham1995} to take into account the energy loss
in the vertical direction by the standard form
\begin{equation}
  F_z = \frac{4\sigma_\mathrm{SB} T^4}{3\tau_\mathrm{d}},
\end{equation}
where we use $\sigma_\mathrm{SB}$ to denote the Stefan-Boltzmann constant. The optical depth $\tau_\mathrm{d}$ is again
chosen based on the physical problem. We finally set to zero the other two terms, $\nabla\cdot\mathbf{q}$ and
$\nabla\cdot\mathbf{F}$, in the energy equation.

%---------------------------------------------------------------------------------------------------------------------

\section{Numerical Method}

In our numerical algorithm, we use pseudo-spectral methods to evaluate the spatial derivatives in the partial
differential equations and an explicit Runge-Kutta method to advance the solution in time.

Spectral methods are based on the idea that any function can be expanded in a series of orthogonal basis functions.
When a well-behaved basis is chosen, this series converges exponentially in the absence of discontinuities and the
numerical partial derivatives can be easily evaluated. There exist different ways to approximate functions by a series
with a finite number of terms, such as the Galerkin method, the tau method, and the pseudo-spectral method
\citep[see][]{Canuto1988,Gottlieb1983,Guo1998,Boyd2000,Peyret2002}. In pseudo-spectral methods, as the one we use here,
we choose a set of collocation points and then evaluate the expansion (spectral) coefficients such that the truncated
series agree exactly with the original functions at the collocation points up to the machine accuracy. In the next
subsection, we discuss our implementation of the pseudo-spectral method.

\subsection{Collocation Methods}\label{sec:collocation}

The Fourier basis is the natural choice for expanding functions with periodic boundary conditions, as is the case along
the azimuthal direction in our problem. Any physical quantity $f = f(r, \phi)$ can be expanded in this basis as
\begin{equation}
  f(r, \phi) = \sum^\infty_{m=-\infty}\hat f_m(r)e^{im\phi}. \label{eq:fourier_expansion}
\end{equation}
We use a total of $M$ evenly spaced, discrete collocation points in the domain $[-\pi, \pi)$,
\begin{equation}
  \phi_j \equiv \frac{2\pi j}{M},\ \ \ \mbox{for}\ -\frac{M}{2} \le j \le \frac{M}{2}-1,
\end{equation}
so that the above Fourier series can be approximated by the (truncated) discrete Fourier series
\begin{equation}
  f(r, \phi_j) = \sum^{M/2-1}_{m=-M/2}\hat f_m(r)e^{i2\pi mj/M}.
\end{equation}
The derivative of $f$ with respect to $\phi$ is, therefore,
\begin{equation}
  \left.\frac{\partial f}{\partial\phi}\right|_{\phi_j} = \sum^{M/2-1}_{m=-M/2}im\hat f_m(r)e^{i2\pi mj/M}.
  \label{eq:fourier_derivative}
\end{equation}
In order to evaluate equation~(\ref{eq:fourier_derivative}), we use a fast Fourier transform algorithm to find $\hat
f_m$, multiply the Fourier coefficients by $im$, and then take the inverse fast Fourier transform of the series. The
computational order of the azimuthal derivative is therefore $\mathcal{O}(M\log_2 M)$.

When non-periodic boundary conditions are present, as in the case along the radial direction, the Fourier basis can no
longer be used. Popular choices of bases are the Legendre and Chebyshev polynomials, which are both defined on the
domain $[-1,1]$. Between these two, the Chebyshev polynomials are usually preferred because of their relation to the
cosine function, which makes them easier to compute. Suppose the domain of $r$ is $[r_{\min}, r_{\max}]$. In order to
map the standardized radial coordinate $\bar r \in [-1,1]$ to the physical radial coordinate $r \in [r_{\min},
r_{\max}]$, we introduce a mapping function $r = g(\bar r)$ which is strictly increasing and satisfies both $g(-1) =
r_{\min}$ and $g(1) = r_{\max}$. Let $T_n(\bar r)$ be the $n$-th order Chebyshev polynomial, i.e.,
\begin{equation}
  T_n(\bar r) \equiv \cos\left(n\arccos\bar r\right).
\end{equation}
The Chebyshev-Gauss-Lobatto collocation points are defined by
\begin{equation}
  \bar r_k \equiv \cos\left(\frac{\pi k}{N}\right),\ \ \ \mbox{for $0 \le j \le N$},
\end{equation}
where the collocation points in the physical radial coordinate are given by $r_k = g(\bar r_k)$. Note that there are
totally $N+1$ collocation points including both boundary points of the domain. For any physical quantity $f(r, \phi)$,
we approximate its Chebyshev-Fourier series by the double sum
\begin{eqnarray}
  f(r_k, \phi_j) & = & f[g(\bar r_k), \phi_j] \nonumber\\
  & = & \sum^{M/2-1}_{m=-M/2}\sum^N_{n=0}\check f_{n,m}T_n(\bar r_k)e^{i2\pi mj/M} \nonumber\\
  & = & \sum^{M/2-1}_{m=-M/2}\sum^N_{n=0}\check f_{n,m}\cos\left(\frac{\pi nk}{N}\right)e^{i2\pi mj/M}.
\end{eqnarray}
We then use a fast discrete cosine transform to obtain the spectral coefficient $\check f_{n,m}$ from $\hat f_{m}(r) =
\hat f_{m}[g(\bar r)]$. The radial derivative can be found using the chain rule
\begin{equation}
  \frac{\partial f}{\partial r} = \frac{1}{dg/d\bar r}\frac{\partial f}{\partial\bar r}.
\end{equation}
Expressing the derivative with respect to the standardized coordinate as
\begin{equation}
  \frac{\partial f}{\partial \bar r} = \sum^{M/2-1}_{m=-M/2}\sum^N_{n=0}\check f_{n,m}^{(1)}T_n(\bar r)e^{im\phi},
  \label{eq:chebyshev_derivative}
\end{equation}
we can employ the following three-term recursive relation to obtain $\check f_{n,m}^{(1)}$ from $\check f_{n,m}$,
\begin{equation}
  \left\{\begin{array}{rcl}
    \check f_{N,m}^{(1)} & = & 0, \\
    \check f_{N-1, m}^{(1)} & = & 2N\check f_{N,m}, \\
    c_n\check f_{n,m}^{(1)} & = & \check f_{n+2,m}^{(1)} + 2(n+1)\check f_{n+1,m},
  \end{array}\right.
\end{equation}
where $c_0 = 2$, and $c_n = 1$, for $n>1$. This recursion relation allows the use of an $\mathcal{O}(N)$ algorithm in
calculating the numerical derivatives in spectral space along the radial direction. The transformation between physical
and spectral space can then be done by a fast cosine transform, the order of which is $\mathcal{O}(N\log_2 N)$.

The most trivial map $r = g(\bar r)$ is linear, i.e.,
\begin{equation}
  r = \frac{r_{\max}}{2}(\bar r + 1) - \frac{r_{\min}}{2}(\bar r - 1).
\end{equation}
In this case,
\begin{equation}
  \frac{1}{dg/d\bar r} = \frac{d\bar r}{dr} = \frac{2}{r_{\max} - r_{\min}}
  \label{eq:chebysehv_derivative_linear_scaling}
\end{equation}
is just a constant. However, the grid spacings $\Delta r_k = r_k - r_{k-1}$, for $1 \le k \le N$, scale with the number
of collocation points $N$ as $\Delta r_k = \mathcal{O}(N^{-2})$, when $k$ is close to $1$ or $N$. This requires the
stability condition for time-stepping to be
\begin{equation}
  \Delta t = \mathcal{O}(N^{-4})
\end{equation}
due to viscous time scale (see \S\ref{sec:time_stepping}). The grid spacings for Fourier collocation are $\Delta\phi_j
= \phi_j - \phi_{j-1}$. They are uniform and scale as $\Delta\phi_j = \mathcal{O}(M^{-1})$ which gives the stability
condition
\begin{equation}
  \Delta t = \mathcal{O}(M^{-2}).
\end{equation}
The stability condition in the radial direction is too expensive numerically compared to the one required by the
azimuthal direction. To overcome this restriction, \citet{Kosloff1993} proposed the mapping
\begin{equation}
  r = \frac{r_{\max}}{2}\left[\frac{\arcsin(\alpha \bar r)}{\arcsin(\alpha)} + 1\right]
    - \frac{r_{\min}}{2}\left[\frac{\arcsin(\alpha \bar r)}{\arcsin(\alpha)} - 1\right],
    \label{eq:modified_chebyshev_collocation}
\end{equation}
which makes the spacing $\Delta r_k = \mathcal{O}(N^{-1})$ around the boundaries. In the above transformation, $\alpha$
is a parameter. Let $\epsilon$ be the machine accuracy. \citet{Don1997} showed that the choice
\begin{equation}
  \alpha = \mathrm{sech}\left(\frac{|\ln \epsilon|}{N}\right) \label{eq:optimized_alpha}
\end{equation}
minimizes the errors in the numerical derivatives. We therefore refer to the
mapping~(\ref{eq:modified_chebyshev_collocation}) with the optimized choice (\ref{eq:optimized_alpha}) as the modified
Chebyshev collocation and employ it in our algorithm. The numerical derivatives for the modified Chebyshev collocation
can be easily calculated by
\begin{equation}
  \left.\frac{\partial f}{\partial r}\right|_{r_k}
  = \frac{2\arcsin(\alpha)}{(r_{\max} - r_{\min})\alpha}\sqrt{1-(\alpha r_k)^2}
  \left.\frac{\partial f}{\partial \bar r}\right|_{\bar r_k},
  \label{eq:chebysehv_derivative_scaling}
\end{equation}
where $(\partial f/\partial \bar r)|_{\bar r_k}$ is obtained by the usual Chebyshev
method~(\ref{eq:chebyshev_derivative}). The constant scaling in (\ref{eq:chebysehv_derivative_linear_scaling}) and the
scaling in equation (\ref{eq:chebysehv_derivative_scaling}) can be done together with the normalization, so neither of
them have extra computational cost.

\subsection{Treatment of the Non-Linear Terms}\label{sec:non-linear}

In low order numerical methods, one uses the conservative form of the hydrodynamic equations to ensure conservation of
mass and momentum \citep{LeVeque1992}. This is not necessarily the ideal way for pseudo-spectral methods because the
conservative form is sometime numerically unstable \citep[see][pp.286--289]{Peyret2002}. To illustrate this, consider
two arbitrary functions $u(x)$ and $v(x)$. The derivative of their product has two different but equivalent forms
\begin{eqnarray}
  C & = & (uv)',\\
  D & = & u'v + uv'.
\end{eqnarray}

The pseudo-spectral Fourier coefficients (with $M$ collocation points) of $C$ and $D$ are
\begin{eqnarray}
  \hat C_m & = & im\sum_{p+q = m}\hat u_p\hat v_q + i\left[\sum_{p+q = m+M} m\hat u_p\hat v_q \right.\nonumber\\
  & & \left.+ \sum_{p+q = m-M}m\hat u_p\hat v_q\right], \\
  \hat D_m & = & im\sum_{p+q = m}\hat u_p\hat v_q + i\left[\sum_{p+q = m+M}(m+M)\hat u_p\hat v_q \right.\nonumber\\
  & & \left.+ \sum_{p+q = m-M}(m-M)\hat u_p\hat v_q\right].
\end{eqnarray}
The first term in each equation corresponds to the truncated Fourier series. The other terms in the square brackets are
therefore due to the aliasing error of pseudo-spectral methods \citep[see][pp.202--221]{Boyd2000}, which may cause
numerical instability. One can show that these aliasing terms have opposite signs. The skew symmetric form $(\hat C_m +
\hat D_m)/2$, therefore, has a much smaller aliasing error. Unfortunately, calculating the skew symmetric form
increases the number of numerical derivatives and lowers the algorithmic performance.

A similar argument is valid for the Chebyshev collocation. However, in that case, the problem becomes more complicated
due to the boundary conditions. \citet{Botella2001} carried out numerical experiments and reported that, without
aliasing removal, the convective form is stable for the two different boundary method they tested; the conservative
form is unstable for one of their methods. We carried out some numerical experiments for our algorithm. When the
initial conditions are smooth, both conservative and convective forms are stable. However, when an initial perturbation
is introduced, the conservative form sometimes become unstable. In these cases, the spectral coefficients at high
wavenumbers in the radial direction grow exponentially even with spectral filtering (see the next subsection). The
convective form, on the other hand, is able to reproduce various analytical stability criteria, as shown in
\S\ref{sec:tests}.

Because the convective form gives an approximation that is of the same order as the conservative form, in the present
implementations of the numerical algorithm, we use the convective form for the non-linear terms in Navier-Stokes
equation to ensure numerical stability. The continuity and energy equations, on the other hand, are written in
conservative form to conserve mass and energy. The exact equations we solve in the algorithm can be found in
appendix~\ref{app:equations}.

\subsection{Spectral Filtering}\label{sec:spectral_filter}

In our numerical algorithm, we use a combination of spectral filters in order to resolve the two principle drawbacks of
spectral methods. First, because of the non-linear character of the Navier-Stokes equations described above, we need to
filter out the high-frequency modes in each time step to reduce the aliasing error. The spectral filters ensure
long-time stability of the algorithm. Second, when shocks are present in the solutions, the spectral coefficients do
not converge exponentially and oscillations appear around the discontinuities (Gibbs phenomenon). Introducing an
additional spectral filter increases the converge rate of the solutions. Indeed, it has been proven that the filtered
solution converges to the correct entry solution and hence gives the correct shock properties. \citep[For more
information see][and reference therein.]{Don1994,Don1995,Gottlieb1997,Ma1998a, Ma1998b,Guo2001,Li2001}

Suppose we want to apply a filter in the $\phi$-direction. We first find the Fourier coefficients $\hat f_m(r)$
according to equation~(\ref{eq:fourier_expansion}). We then denote the filtered sum by
\begin{equation}
  f^{\sigma\!_\beta}(r, \phi_j) = \sum^{M/2-1}_{m=-M/2}\sigma\!_\beta\left(\frac{2m}{M}\right)\hat f_m(r)e^{i2\pi mj/M},
\end{equation}
where we use the exponential filter
\begin{equation}
  \sigma\!_\beta\left(\frac{2m}{M}\right) = \exp\left(-|\ln\epsilon|\left|\frac{2m}{M}\right|^{\beta}\right).
\end{equation}
The parameter $\epsilon$ here is again the machine accuracy. We choose $\beta \approx M/2$ so that the filtered sum
approximates the original function very well and does not reduce the accuracy of the numerical solution. The same
filter is used in the radial direction, i.e.,
\begin{equation}
  \sigma\!_\beta\left(\frac{n}{N}\right) = \exp\left(-|\ln\epsilon|\left|\frac{n}{N}\right|^{\beta}\right),
\end{equation}
where $\beta$ is chosen close to $N$ so the double filtered sum is
\begin{eqnarray}
  f^{\sigma\!_{N},\sigma\!_{M/2}}(r_k, \phi_j) & = & \sum^{M/2-1}_{m=-M/2}\sum^N_{n=0}
  \sigma\!_N\left(\frac{n}{N}\right)\sigma\!_{M/2}\left(\frac{2m}{M}\right)\nonumber\\
  & & \ \ \ \times\check f_{n,m}\cos\left(\frac{\pi nk}{N}\right)e^{i2\pi mj/M}.
\end{eqnarray}

\subsection{Initial Perturbations}\label{sec:perturbation}

\begin{figure*}
  \plottwo{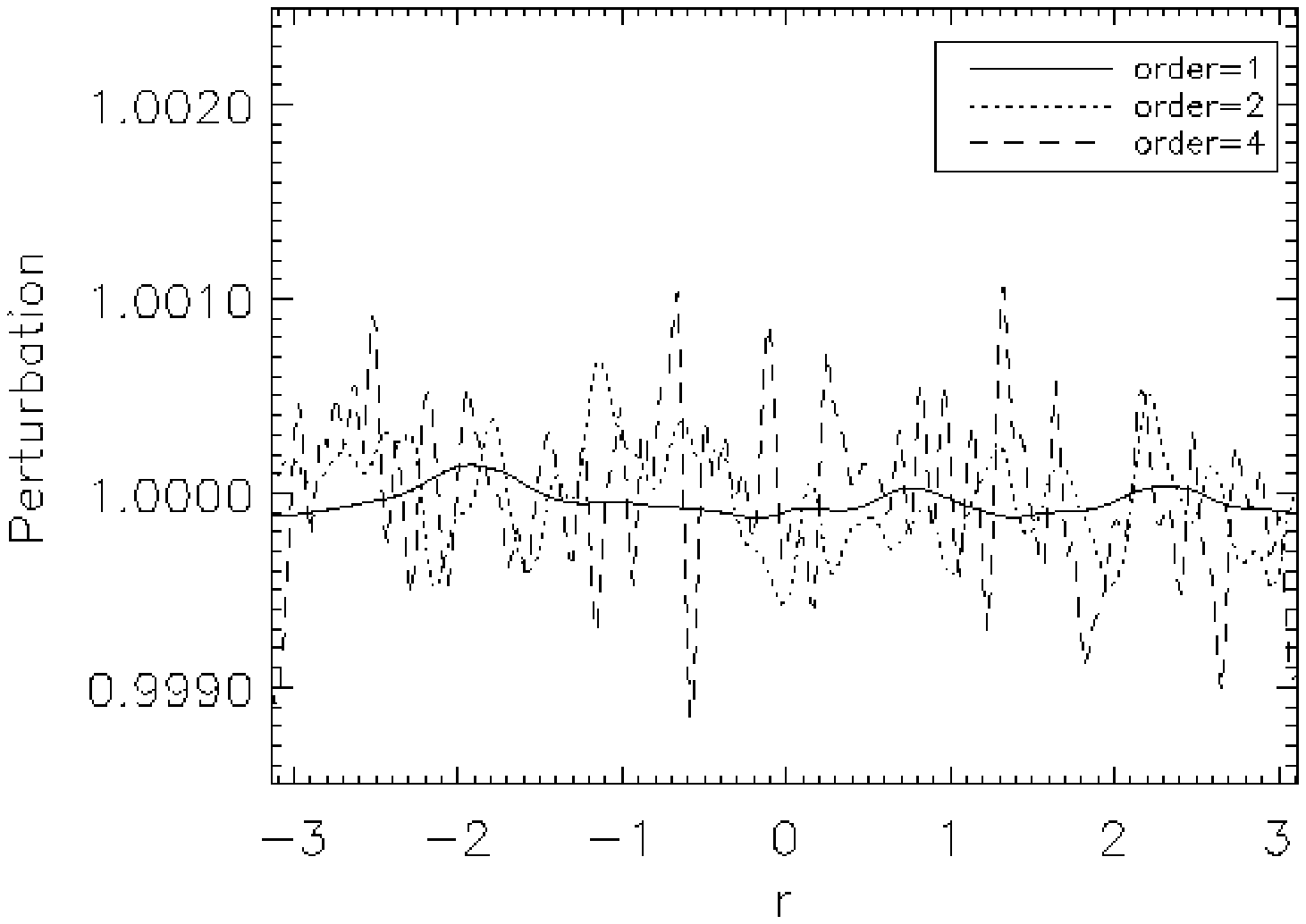}{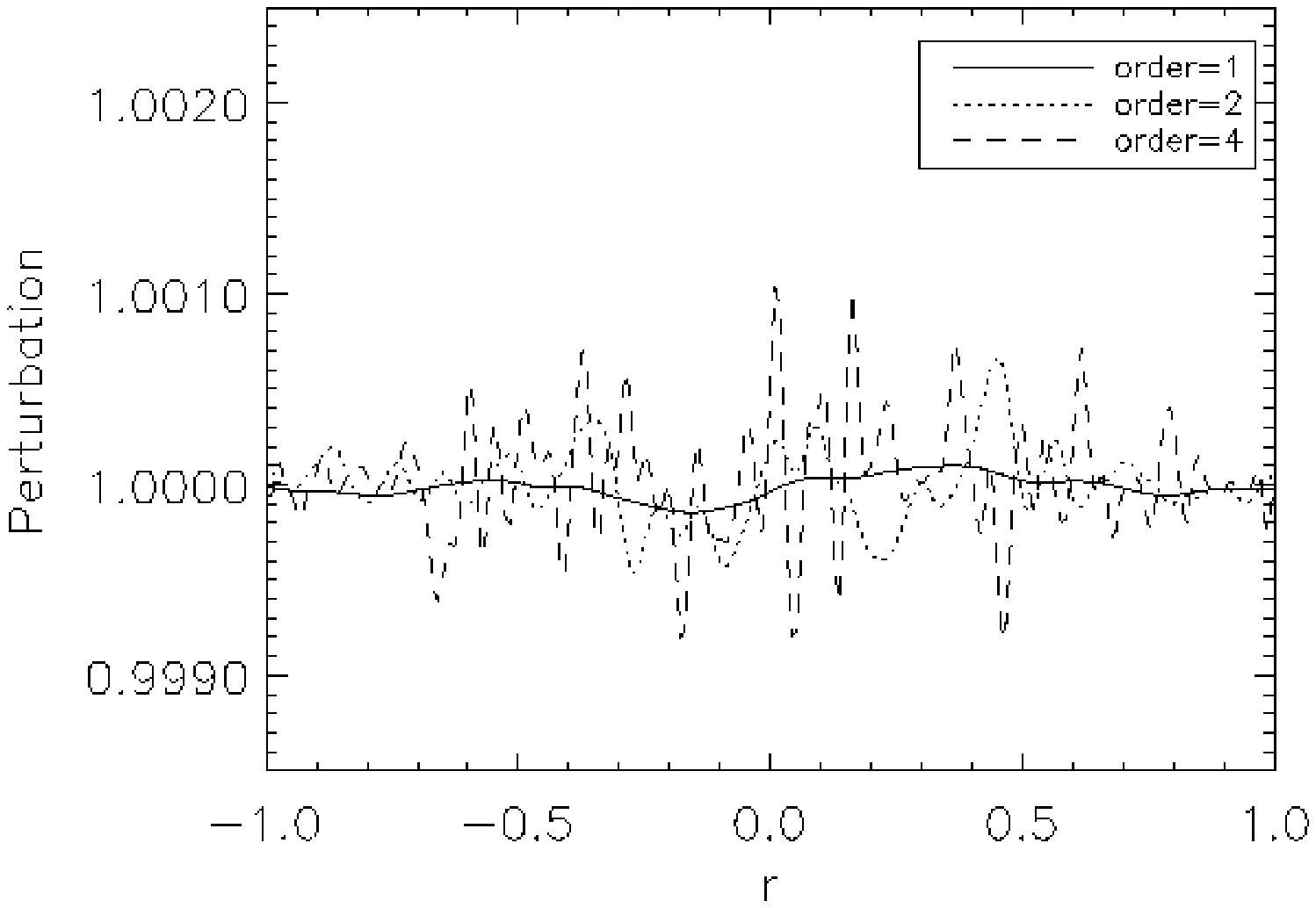}
  \caption{Two 1-dimension perturbation profiles generated by the method describe in \S\ref{sec:perturbation}. All
           lines in the plots have $\gamma_0 = 0.001$. \emph{(Left)} The function is assumed periodic and approximated
           by a Fourier collocation. \emph{(Right)} A Chebyshev collocation is used to demonstrate the behavior of our
           method near the boundaries.}
           \label{fig:perturbation}
\end{figure*}

Spectral methods have been used successfully to study non-randomized perturbation, such as the tidal effects presented
by \citet{Godon1997,Godon1998}. Later, \citet{Godon2000} described a method to add random perturbations to their
initial conditions. In their simulation, they generated 100 Gaussian profiles randomly within the computation domain.
This method ensure that the spectral coefficients of the perturbed function converge exponentially, and avoids changing
the direction of the characteristic at the boundary.

In our calculations, we require the initial perturbations to be smooth, i.e., we require the spectrum of the
perturbations to converge exponentially. One simple method to obtain such a perturbation is to generate random noise in
spectral space, multiply the noise by an exponential filter, and transform the noise back to physical space.

To formulate this method, we first generate a flat-noise spectrum, $\check P_{n,m}$, which is totally random. The
perturbation is added to the original function by the following equation
\begin{equation}
  f(r,\phi) = [1 + \gamma(r,\phi)P^{\sigma\!_\beta,\sigma\!_\beta}(r,\phi)]f_0(r,\phi),
\end{equation}
where $f(r,\phi)$ is the perturbed function, $\gamma(r, \phi)$ is the size of the perturbation,
$P^{\sigma\!_\beta,\sigma\!_\beta}(r,\phi)$ is the filtered sum
\begin{eqnarray}
  P^{\sigma\!_{\beta},\sigma\!_{\beta}}(r_k, \phi_j) & = & \sum^{M/2-1}_{m=-M/2}\sum^N_{n=0}
  \sigma\!_N\left(\frac{n}{\beta}\right)\sigma\!_{M/2}\left(\frac{m}{\beta}\right)\nonumber\\
  & & \ \ \ \times\check P_{n,m}\cos\left(\frac{\pi nk}{N}\right)e^{i2\pi mj/M},
\end{eqnarray}
as defined in the pervious subsection, and $f_0(r,\phi)$ is the unperturbed function. The order of filter, $\beta$,
controls the spacial scale of the perturbation. Because $f_0(r, \phi)$ is periodic in $\phi$, we can simply use a
$\phi$-independent function $\gamma(r, \phi) = \gamma(r)$. However, in the $r$-direction we do not want the
perturbation to extend the boundaries. We, therefore, set
\begin{equation}
  \gamma(r) = \gamma_0 \frac{\arcsin(\alpha)}{\alpha}\sqrt{1 - (\alpha r)^2},
\end{equation}
which is pre-computed by equation~(\ref{eq:chebysehv_derivative_scaling}), when we use the modified Chebyshev
collocation method.

In Figure~\ref{fig:perturbation}, we illustrate this method of introducing perturbations to an one-dimensional constant
function $f_0 = 1$. The left plot uses a Fourier basis with 256 collocation points. Where the right plot uses a
Chebyshev basis with 257 collocation points. All the lines have $\gamma_0 = 0.001$ but different orders $\beta$. Note
that the extra rescaling for Chebyshev collocation lowers the amplitude of the perturbations around the boundaries.

\subsection{Boundary Conditions}

Solving partial differential equations in a finite domain usually requires Dirichlet or/and Neumann boundary
conditions. When using finite difference methods, one can simply change the values of physical quantities at the
boundary points to achieve Dirichlet boundary conditions. Alternatively, one can adjust the ``ghost points'' to achieve
Neumann boundary conditions. Unfortunately, these methods do not generally work for spectral methods. Indeed, changing
the boundary points (or ghost points) effectively introduces step functions. When discontinuities are present, the
spectral coefficients do not converge exponentially and the spectral methods fail to produce stable solutions.

Perturbations around the boundaries can also change the characteristic directions of the flow. A naive Dirichlet or
Neumann boundary treatment fails to take care of these oscillations between inflowing and outgoing characteristics.
When the inflowing boundary conditions are not given, or when the outgoing boundary conditions are not consistent with
the solutions in the computation domain, the system of differential equations is ill-posed and the algorithm diverges.
Although we generate the initial perturbations in a way that they are small around the boundaries (see the previous
subsection), they can still propagate and reach the boundaries as time evolves.

In our numerical algorithm, we introduce a new boundary treatment in the radial direction, which is a spatial filter.
This method forces each dynamic variable to approach smoothly to its boundary value so the characteristics directions
at the boundaries are well-defined. This ensures that any instabilities of the solution are not due to boundary
conditions. For example, consider a one-dimensional problem
\begin{equation}
  \frac{\partial f(r,t)}{\partial t} = F[f(r,t)]
\end{equation}
for any functional form $F$. If we want to apply a Dirichlet boundary condition at the outer boundary, we choose some
smooth monotonic function $h(r)$ such that $h(r) \rightarrow 1$ for interior points of the computational domain and
$h(r) \rightarrow 0$ at the boundary. At each time step, we impose the boundary condition by the spatial filter
\begin{equation}
  f_k^i \longrightarrow \left(f_k^i - f_0\right)h(r_k) + f_0, \label{eq:boundary_condition}
\end{equation}
where $f_k^i$ denotes the numerical approximation of our function at time $i$ and radial collocation point $k$. This
filter makes the function $f$ approach its boundary value $f_0$ and produces a numerical boundary layer. Rearranging,
the above step is equivalent to setting
\begin{equation}
  f_k^i \longrightarrow f_k^i - [1 - h(r_k)](f_k^i - f_0).
\end{equation}
Hence, it is equivalent to adding an extra source/sink term $\dot f_\mathrm{s}$ in the original equation
\begin{equation}
  \frac{\partial f}{\partial t} = F[f] - \frac{1 - h(r)}{\Delta t}(f - f_0) \equiv F[f] + \dot f_\mathrm{s},
\end{equation}
where $\Delta t$ is the time step (see \S\ref{sec:time_stepping}). Omitting $F$ in the above equation, it becomes
$\partial f/\partial \sim -(f - f_0)$. The value of $[1 - h(r)]/\Delta t$ controls the converge rate of $f$ to $f_0$,
which is automatically proportional to the velocity, because $1/\Delta t \sim |\mathbf{v}|$.

We use an exponential filter that is similar to the spectral filter discussed in \S3.3. To impose an outer boundary
condition, we need the monotonically decreasing filter
\begin{equation}
  h(r) = \sigma\!_\beta\left(\frac{r - r_{\min}}{r_{\max} - r_{\min}}\right)
  = \exp\left[-|\ln\epsilon|\left(\frac{r - r_{\min}}{r_{\max} - r_{\min}}\right)^\beta\right].
\end{equation}
On the other hand, for the inner boundary, we use the monotonically increasing filter
\begin{equation}
  h(r) = \sigma\!_\beta\left(\frac{r_{\max} - r}{r_{\max} - r_{\min}}\right)
  = \exp\left[-|\ln\epsilon|\left(\frac{r_{\max} - r}{r_{\max} - r_{\min}}\right)^\beta\right].
\end{equation}
We can change the thickness of the numerical boundary layer by changing the order of the filter, $\beta$.

This boundary treatment turns out to be very convenient in modeling non-reflective boundary conditions. When we choose
the parameter $\beta$ to be equal to the number of collocation points, the boundary layer is thick enough to discard
any outgoing waves but thin enough so that the interior solution is not affected significantly.

For applying Neumann boundary conditions, the addition of a ghost zone is not practical in spectral methods. This is
because Chebyshev polynomials approximate the derivatives based on all the interior points. \citet{Godon1997} presented
a method which involves solving for the boundary values of physical quantities, so that their derivatives agree with
the boundary conditions. This is effectively a Dirichlet boundary treatment with the boundary values depending on all
the interior values at each time step.

To avoid numerical instabilities due to the Neumann boundary conditions, $(\partial f/\partial r)|_\mathrm{boundary} =
f'_0$, we choose another method that involves applying a spatial filter to each variable after taking the radial
derivatives. This is equivalent to replacing the radial derivatives $(\partial f/\partial r)|_{r_k}$ by
\begin{equation}
  \left.\frac{\partial f}{\partial r}\right|_{r_k}^{\ i}
  \longrightarrow \left.\frac{\partial f}{\partial r}\right|_{r_k}^{\ i}
  - \left[1 - \sigma\!_\beta\left(\frac{r_k - r_{\min}}{r_{\max} - r_{\min}}\right)\right]
  \left(\left.\frac{\partial f}{\partial r}\right|_{r_k}^{\ i} - f'_0\right)
\end{equation}
for the inner boundary, and
\begin{equation}
  \left.\frac{\partial f}{\partial r}\right|_{r_k}^{\ i}
  \longrightarrow \left.\frac{\partial f}{\partial r}\right|_{r_k}^{\ i}
  - \left[1 - \sigma\!_\beta\left(\frac{r_{\max} - r_k}{r_{\max} - r_{\min}}\right)\right]
  \left(\left.\frac{\partial f}{\partial r}\right|_{r_k}^{\ i} - f'_0\right)
\end{equation}
for the outer boundary. This ensures that the first derivatives around the boundaries are smooth.

\subsection{Time Stepping}\label{sec:time_stepping}

We integrate the hydrodynamics equations with low-storage, explicit Runge-Kutta methods. Let
\begin{equation}
  \mathbf{H} \equiv (H_\Sigma, H_{v_r}, H_{v_\phi}, H_E)^t
\end{equation}
be the right hand sides of hydrodynamic equations~(\ref{eq:hydro_begin}) -- (\ref{eq:hydro_end}) and let
\begin{equation}
  \mathbf{u} \equiv (\Sigma, v_r, v_\phi, E)^t
\end{equation}
be the dynamic variable; the superscript $t$ here stands for transpose. The low-storage explicit Runge-Kutta methods
have the general from
\begin{equation}
  \left\{\begin{array}{rcl}
    \mathbf{u}_0 & = & \mathbf{u}(t_n) \\
    \mathbf{Q}_i & = & A_i \mathbf{Q}_{i-1} + \Delta t \mathbf{H}(u_{i-1}) \\
    \mathbf{u}_i & = & \mathbf{u}_{i-1} + B_i \mathbf{Q}_i,\ \ \ \ \ \ i = 1, ..., s \\
    \mathbf{u}(t_{n+1}) & = & \mathbf{u}_s
  \end{array}\right.,
\end{equation}
where $\Delta t$ is the time step and $A_1 = 0$. All other $A_i$'s and $B_i$'s are coefficients that characterize the
scheme and $\mathbf u_i$ are the intermediate stages of $\mathbf u$. The method to obtain these coefficients can be
found in \citet{Peyret2002}. Depending on the complexity of the problem, we choose between second order or third order
Runge-Kutta, and third order or fourth order Carpenter-Kennedy Runge-Kutta methods. The different algorithms are
compared during the verification tests (see \S\ref{sec:tests_free_fall} and fig.~\ref{fig:error_time} below). The third
order Runge-Kutta scheme, proposed by Williamson \citep[See][p.146]{Peyret2002}, turns out to be the most popular in
spectral methods because of its efficiency and accuracy. For this reason, we present here this scheme explicitly:
\begin{equation}
  \left\{\begin{array}{rcl}
    \mathbf{u}_0 & = & \mathbf{u}(t_n) \\
    \mathbf{Q}_1 & = & \Delta t\ \mathbf{H}(\mathbf{u}_0) \\
    \mathbf{u}_1 & = & \mathbf{u}_0 + \frac{1}{3}\mathbf{Q}_1 \\
    \mathbf{Q}_2 & = & - \frac{5}{9}\mathbf{Q}_1 + \Delta t\ \mathbf{H}(\mathbf{u}_1) \\
    \mathbf{u}_2 & = & \mathbf{u}_1 + \frac{15}{16}\mathbf{Q}_2 \\
    \mathbf{Q}_3 & = & - \frac{153}{128}\mathbf{Q}_2 + \Delta t\ \mathbf{H}(\mathbf{u}_2) \\
    \mathbf{u}_3 & = & \mathbf{u}_2 + \frac{8}{15}\mathbf{Q}_3 \\
    \mathbf{u}(t_{n+1}) & = & \mathbf{u}_3.
  \end{array}\right.
\end{equation}

Recalling the definition of $\Delta r_k$ and $\Delta\phi_j$ in \S\ref{sec:collocation}, we define the
Courant-Friedrich-Lewy (CFL) time step in different directions independently,
\begin{equation}
  \Delta t_\mathrm{CFL,r}    = \min\left(\frac{\Delta r_k   }{c_s + v_r   }\right),\ \ \
  \Delta t_\mathrm{CFL,\phi} = \min\left(\frac{r_k\Delta\phi}{c_s + v_\phi}\right),
\end{equation}
where $c_\mathrm{s} = \sqrt{P/\Sigma}$ is the sound speed. Let $\Delta l \equiv \min(\Delta r_k, r_k\Delta\phi_j)$ be
the minimum grid separation. We also define the viscously restricted time step as
\begin{equation}
  \Delta t_{\nu_\mathrm{s}} = \min\left(\frac{\Delta l^2}{\nu_\mathrm{s}}\right),
\end{equation}
where $\nu_\mathrm{s} = \mu_\mathrm{s}/\Sigma$ is the kinematic viscosity coefficient. Using these definitions, the
maximum allowed time step $\Delta t$ is chosen adaptively by
\begin{equation}
  \Delta t = \delta\min(\Delta t_{\mathrm{CFL},r}, \Delta t_{\mathrm{CFL},\phi}, \Delta t_{\nu_\mathrm{s}}).
\end{equation}
Here $\delta$ is a constant that depends on the physical problem. For example, for a free falling dust ring, the flow
is very stable and $\delta$ could be chosen as large as 29. For most of accretion problems with $v_r \ll v_\phi$,
$\delta$ is of order unity.

%---------------------------------------------------------------------------------------------------------------------

\section{Code Verification}\label{sec:tests}

We have verified our numerical algorithm by comparing our numerical results to a number of test problems with
analytical solutions. These problems are designed to test the implementation of the different terms in the equations,
such as the ones that describe gravity, pressure, and viscosity. In this section, we describe these tests in details
\footnote{The simulation corresponding to these test can be find in the webpage
\texttt{http://www.physics.arizona.edu/\~{}chan/research/\\astro-ph/0406073/}.}.

\subsection{Free Fall of a Dust Ring}\label{sec:tests_free_fall}

\begin{figure*}
  \plottwo{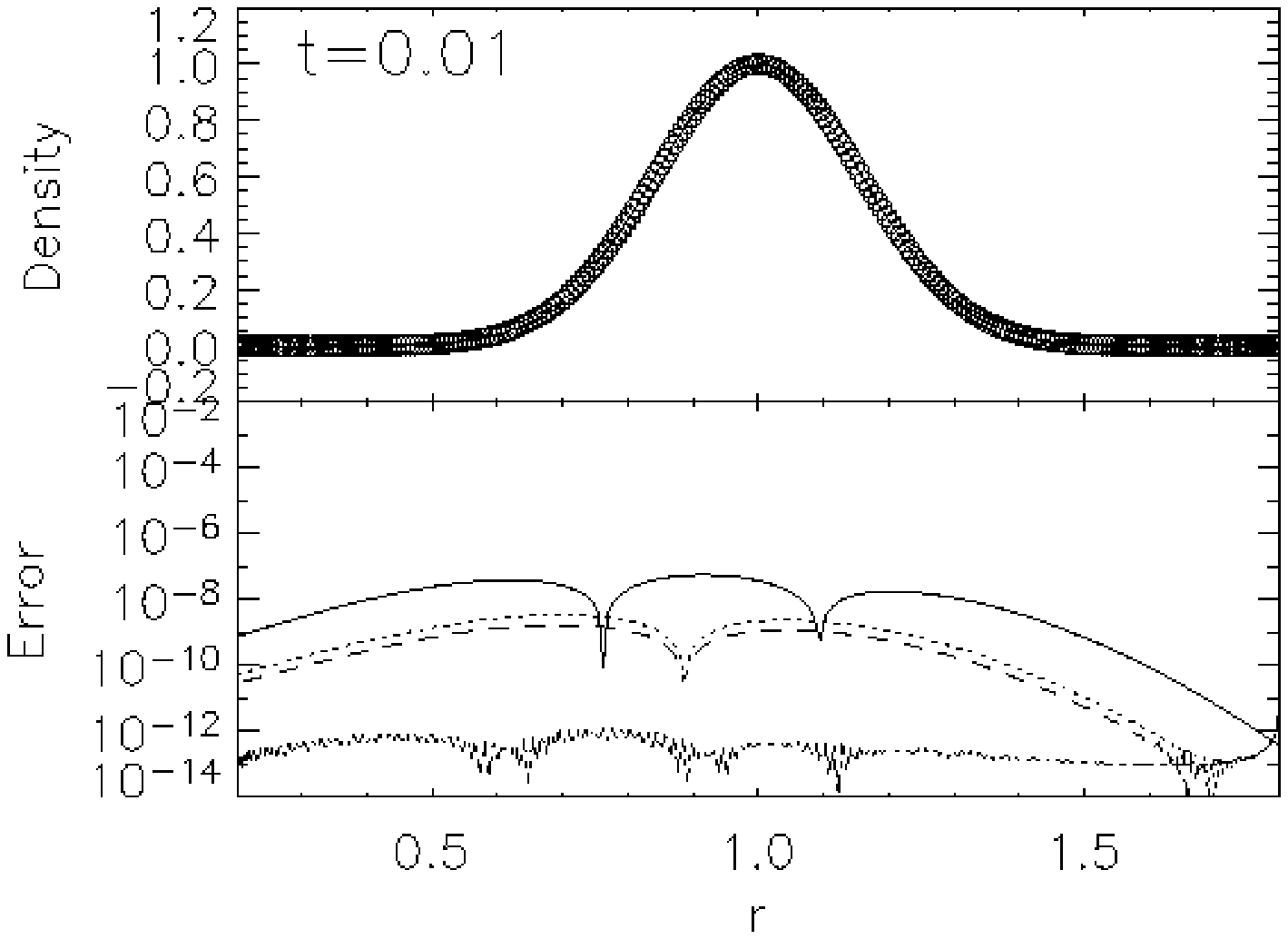}{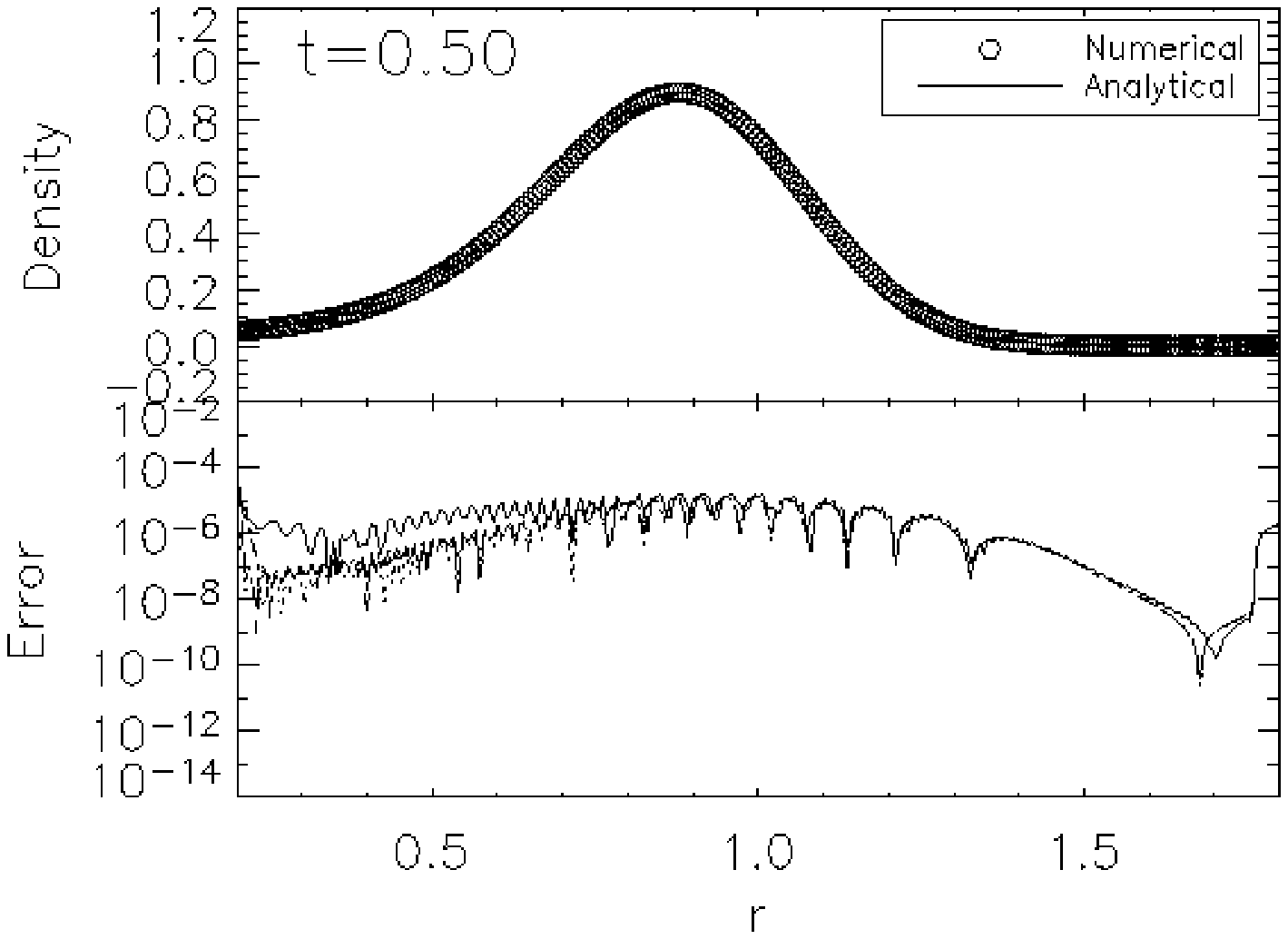} \\
  \plottwo{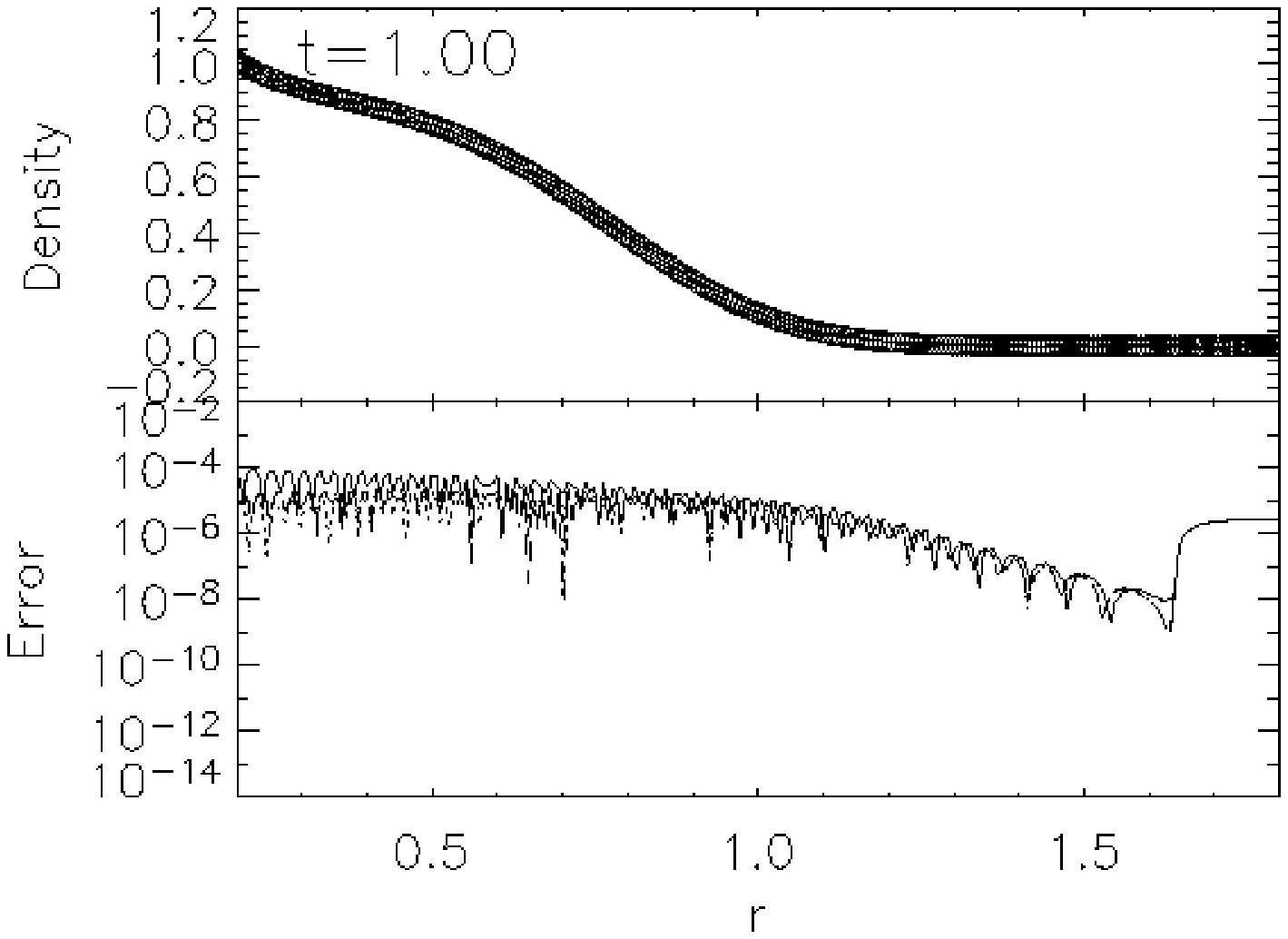}{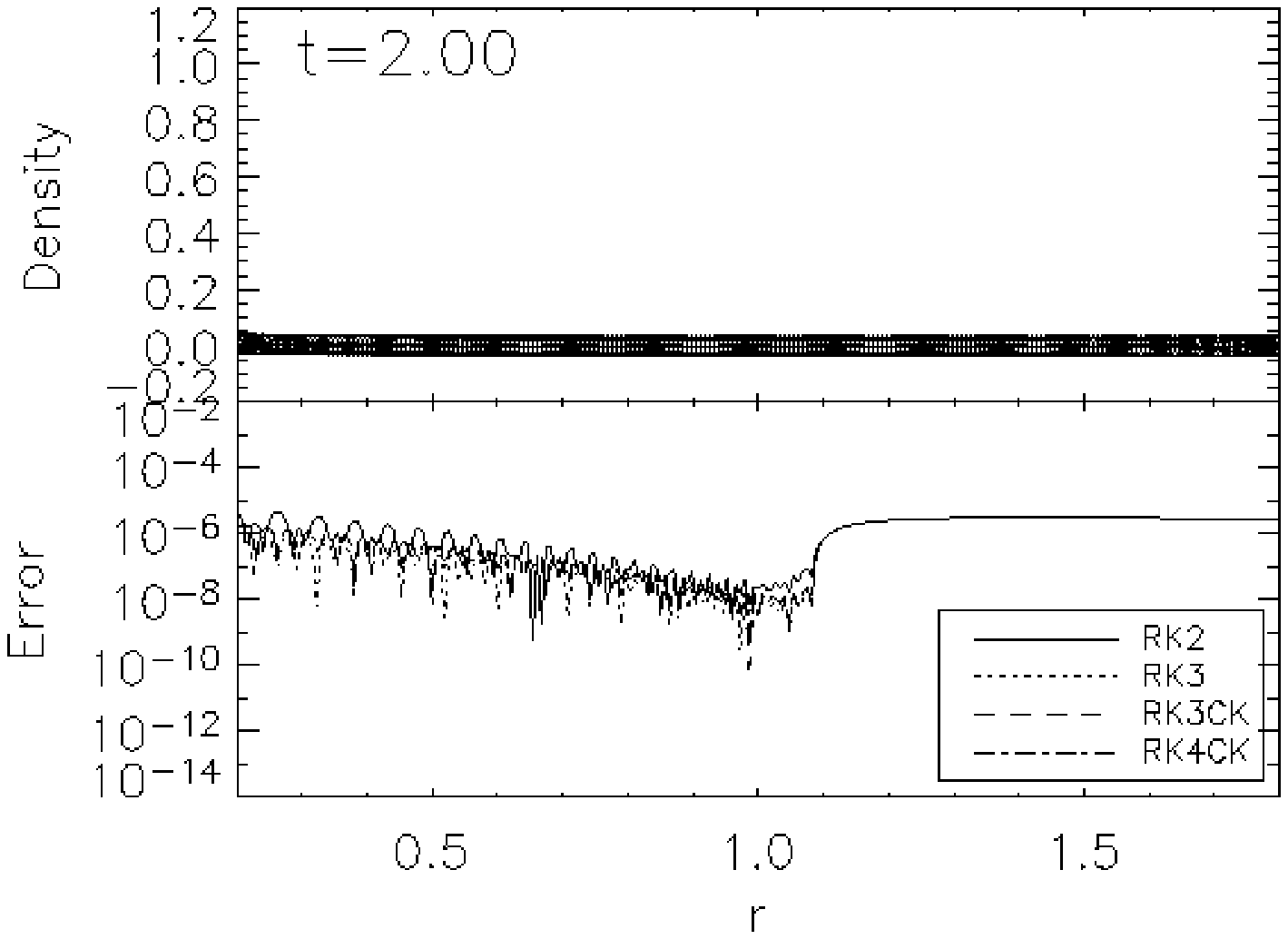}
  \caption{For each panel:
           \emph{(Top)} The analytical solution (solid line) is compared to the numerical solution obtained with a
           third order Runge-Kutta (RK3) method (open circles).
           \emph{(Bottom)} The numerical errors, defined as $|\Sigma_\mathrm{num} - \Sigma_\mathrm{ana}|$, at different
           times for the free-falling Gaussian ring discussed in \S\ref{sec:tests_free_fall}. Different lines
           corresponds to different time-stepping schemes. The legend is in the bottom right plot.}\label{fig:free_fall}
\end{figure*}

\begin{figure*}
  \plottwo{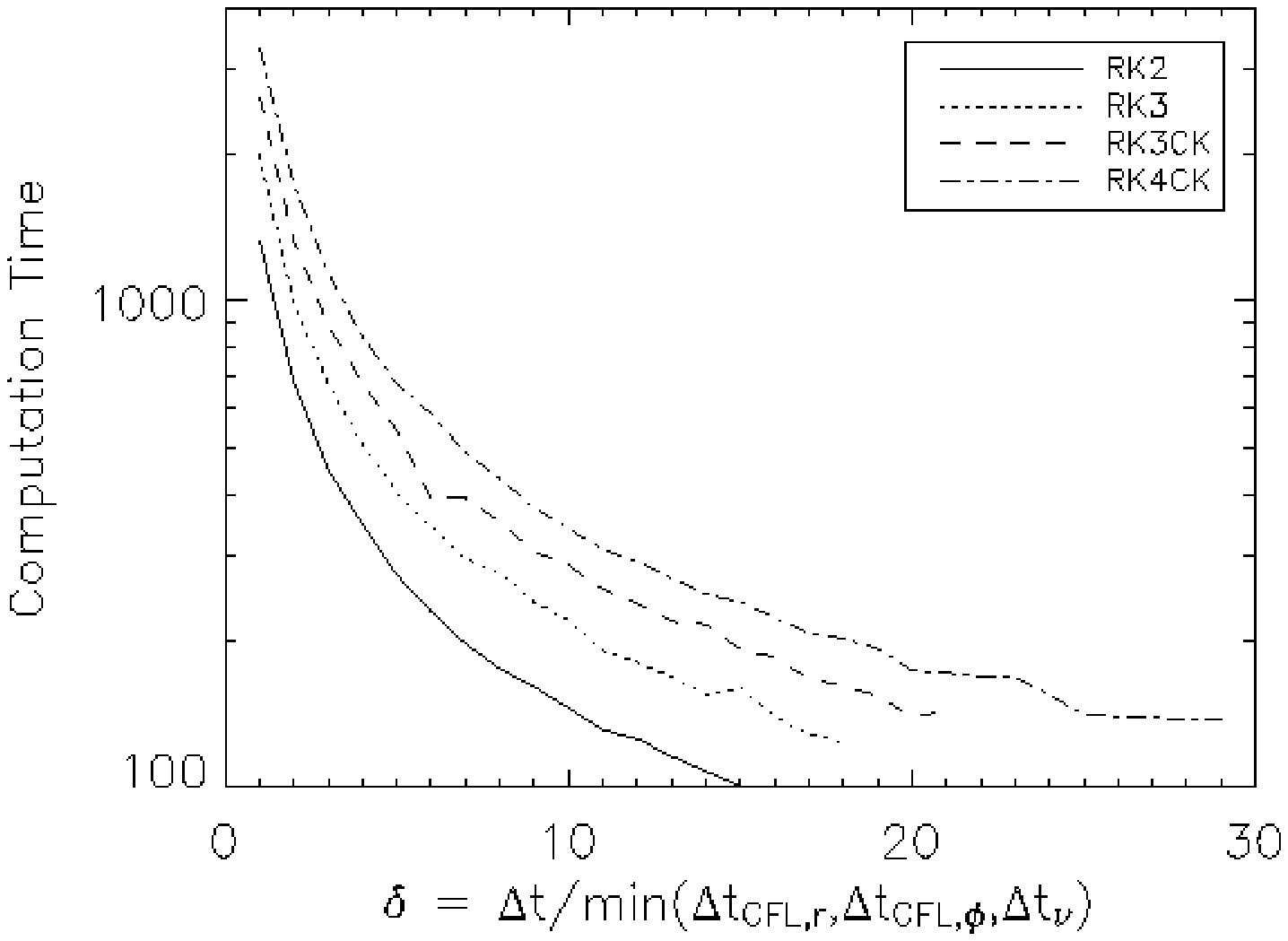}{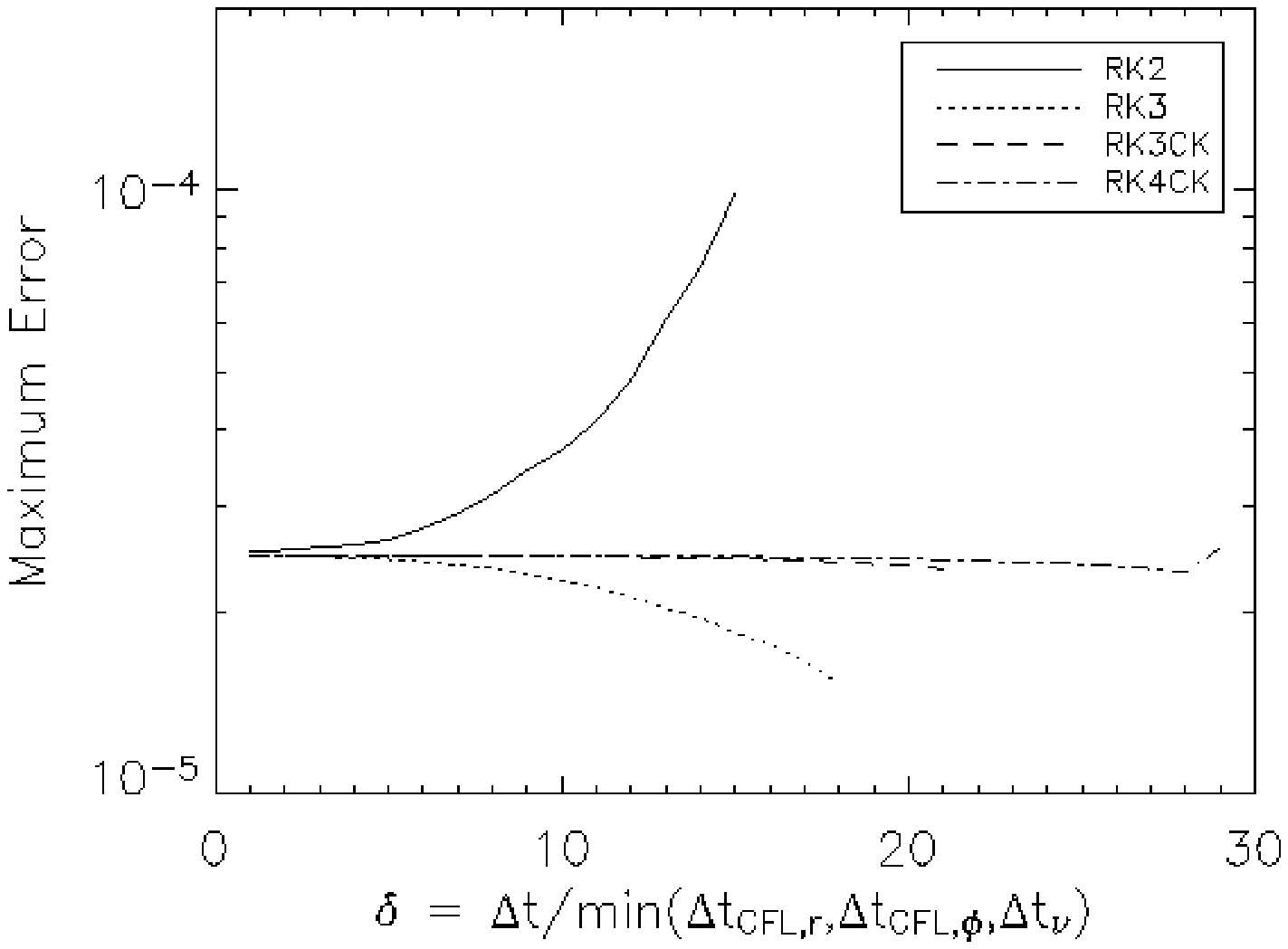}
  \caption{\emph{(Left)} The vertical axis shows the computation time in seconds for the free-falling Gaussian ring
           problem (see \S\ref{sec:tests_free_fall}), with different time stepping schemes.
           \emph{(Right)} The vertical axis shows the numerical errors in the density profile for the same problem. The
           errors in the plot are the maximum of $|\Sigma_\mathrm{num} - \Sigma_\mathrm{ana}|$ at $t = 1$.
           The horizontal axis for both plots are the dimensionless time stepping factor
           $\delta = \Delta t/\min(\Delta t_{\mathrm{CFL},r}, \Delta t_{\mathrm{CFL},\phi}, \Delta t_\nu)$
           (see \S\ref{sec:time_stepping}).}\label{fig:error_time}
\end{figure*}

As a first test, we study the free-fall of an axisymmetric distribution of matter with no angular momentum, pressure,
or viscosity. This problem tests the implementation of gravity, the radial boundary conditions, and the conservation of
mass in the radial direction. Also, we use this problem to evaluate the efficiency and stability of the different
time-stepping schemes. We set $v_\phi = 0$, $P = 0$, $\nu_\mathrm{s} = 0$, and consider only Newtonian gravity with
$\mathbf{g} = (GM/r^2)\mathbf{\hat r}$.

In order to verify our numerical result, we also solve the problem analytically. Using dimensionless quantities so that
$GM = 1$, we write the conservation of energy for a fluid element initially at $r_0$ as
\begin{equation}
  \left(\frac{dr}{dt}\right)^2 = \frac{2}{r} - \frac{2}{r_0}.
\end{equation}
Integrating over time, we obtain the trajectory $r = r(t;r_0)$ of each fluid element in the implicit form
\begin{equation}
  \frac{\sqrt{2}}{r_0\;^{3/2}}t = \frac{1}{2}\sin\left(2\arccos\sqrt{\frac{r}{r_0}}\right)
  + \arccos\sqrt{\frac{r}{r_0}}. \label{eq:r_0}
\end{equation}
We now consider conservation of mass, i.e., at any time we require that
\begin{equation}
  2\pi r\Sigma(r,t)dr = 2\pi r_0\Sigma_0(r_0)dr_0.
\end{equation}
Evaluating $dr_0$ by using equation~(\ref{eq:r_0}), i.e.,
\begin{equation}
  dr_0 = \frac{r_0dr}{\sqrt{r_0/r - 1}}\left[\frac{3}{2}\sqrt{\frac{2}{r_0}}\ t + \frac{r}{\sqrt{r_0/r-1}}\right]^{-1},
\end{equation}
the analytical solution for the density becomes
\begin{equation}
  \Sigma(r,t) = \Sigma_0(r_0)\frac{r_0^2}{r}\left[\frac{3t}{2}
  \sqrt{\left(\frac{2}{r_0}\right)\left(\frac{r_0}{r} - 1\right)} + r\right]^{-1},
\end{equation}
where $r_0$ can be obtained implicitly using equation~(\ref{eq:r_0}).

Although this is an one dimensional problem, we simulate it in the two-dimensional domain $[0.2,1.8]\times[-\pi, \pi]$
with $257\times64$ collocation points. The initial condition is a Gaussian ring
\begin{equation}
  \Sigma_0(r,\phi) = \exp[-20(r-1)^2]
\end{equation}
with zero initial velocity. At the outer boundary, we use the standard Neumann boundary conditions, i.e., we set
$(\partial\Sigma/\partial r)|_{r_\mathrm{out}} = 0$ and $(\partial v_r/\partial r)|_{r_\mathrm{out}} = 0$ for all
times. Moreover, because the characteristics at the inner boundary point towards the negative $r$-direction and we have
neglected viscosity and pressure, we do not need to impose any explicit boundary conditions there. Because filtering
stabilizes the flow, we do not apply any filters in this test in order to compare how different time-stepping methods
affect the stability of the algorithm.

The top panels of Figure~\ref{fig:free_fall} compare the results of the simulations to the analytical solutions; the
open circles denote the numerical results using the third order Runge-Kutta method and the solid lines are the
analytical solution.  The numerical results are indistinguishable from the analytical solution. The bottom panels of
Figure~\ref{fig:free_fall} show the error between the numerical and analytical solution for the maximum stable time
step.

We test four different time-stepping schemes: second order Runge-Kutta (RK2), third order Runge-Kutta (RK3), third
order Carpenter-Kennedy (RK3CK), and four order Carpenter-Kennedy (RK4CK). Figure~\ref{fig:error_time} shows the result
of this test. In the left plot, the vertical axis shows the computation time. The horizontal axis shows the
dimensionless time-stepping factor $\delta = \Delta t/\min(\Delta t_{\mathrm{CFL},r}, \Delta t_{\mathrm{CFL},\phi},
\Delta t_\nu)$ (see \S\ref{sec:time_stepping}). The right plot shows the numerical errors, $\max(|\Sigma_\mathrm{num} -
\Sigma_\mathrm{ana}|)$, at $t = 1$, for the same problem. For $\delta > 15$ in RK2, $\delta
> 18$ in RK3, $\delta > 21$ in RK3CK, and $\delta > 29$ in RK4CK, the solutions diverge so neither the computation time
nor the error are plotted in the graphs. This test demonstrates the ability of our algorithm to conserve mass to an
accuracy better than a few parts in a million. It also demonstrates the advantage of the third order Runge-Kutta method
over other algorithms, when solutions of high accuracy are necessary.

\subsection{Rayleigh's Criterion}\label{sec:tests_rayleigh}

\begin{figure*}
  \plottwo{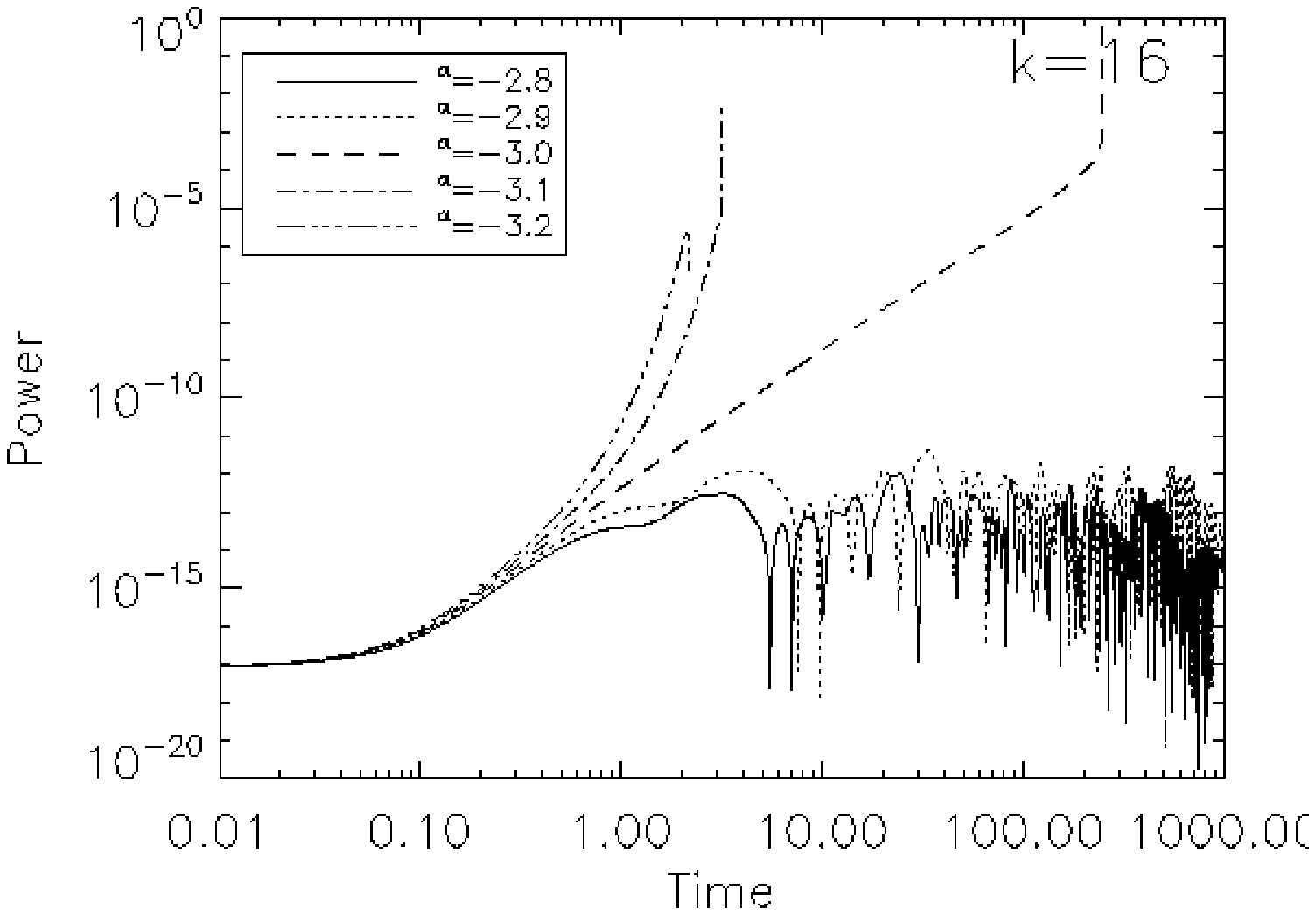}{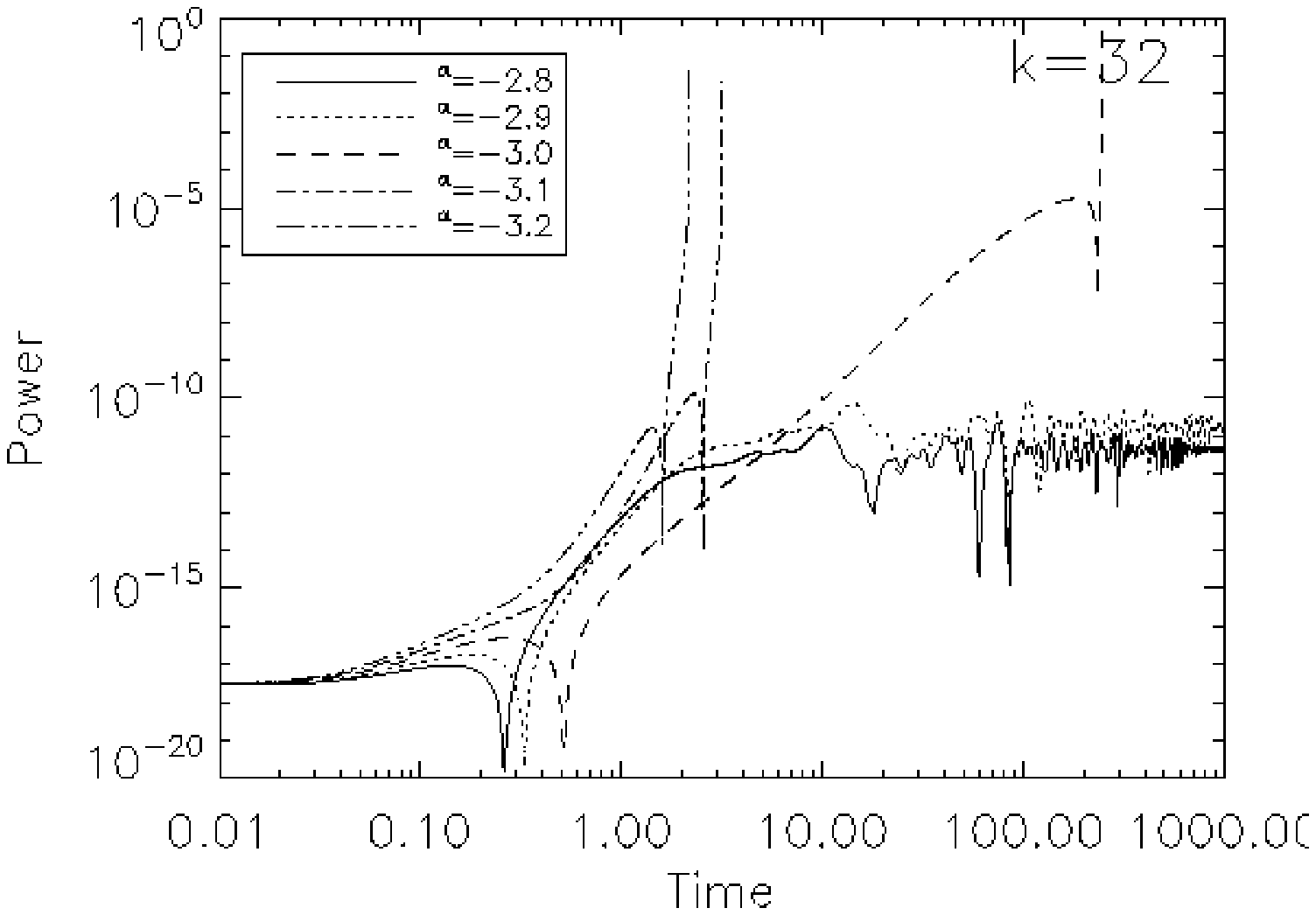} \\
  \plottwo{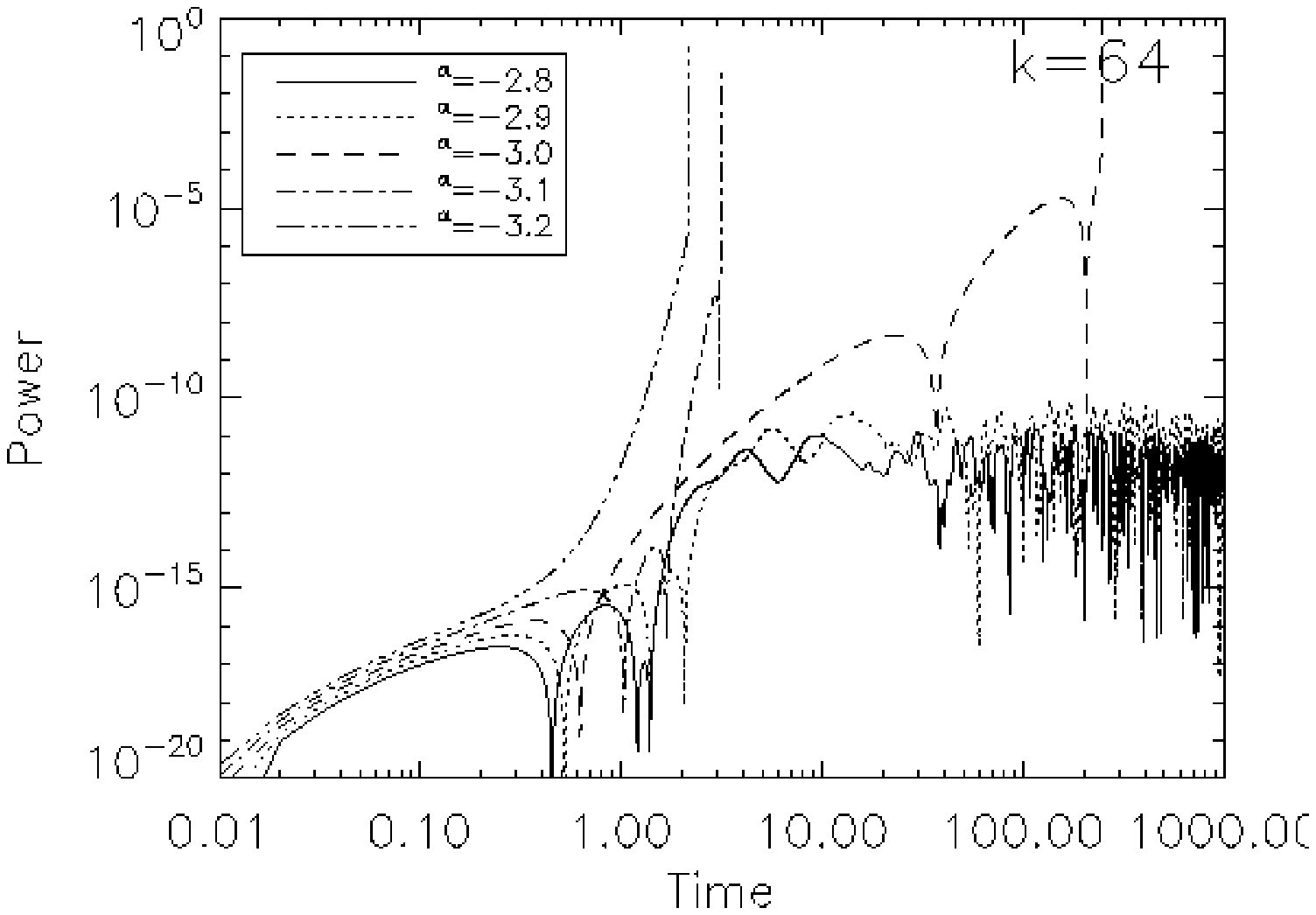}{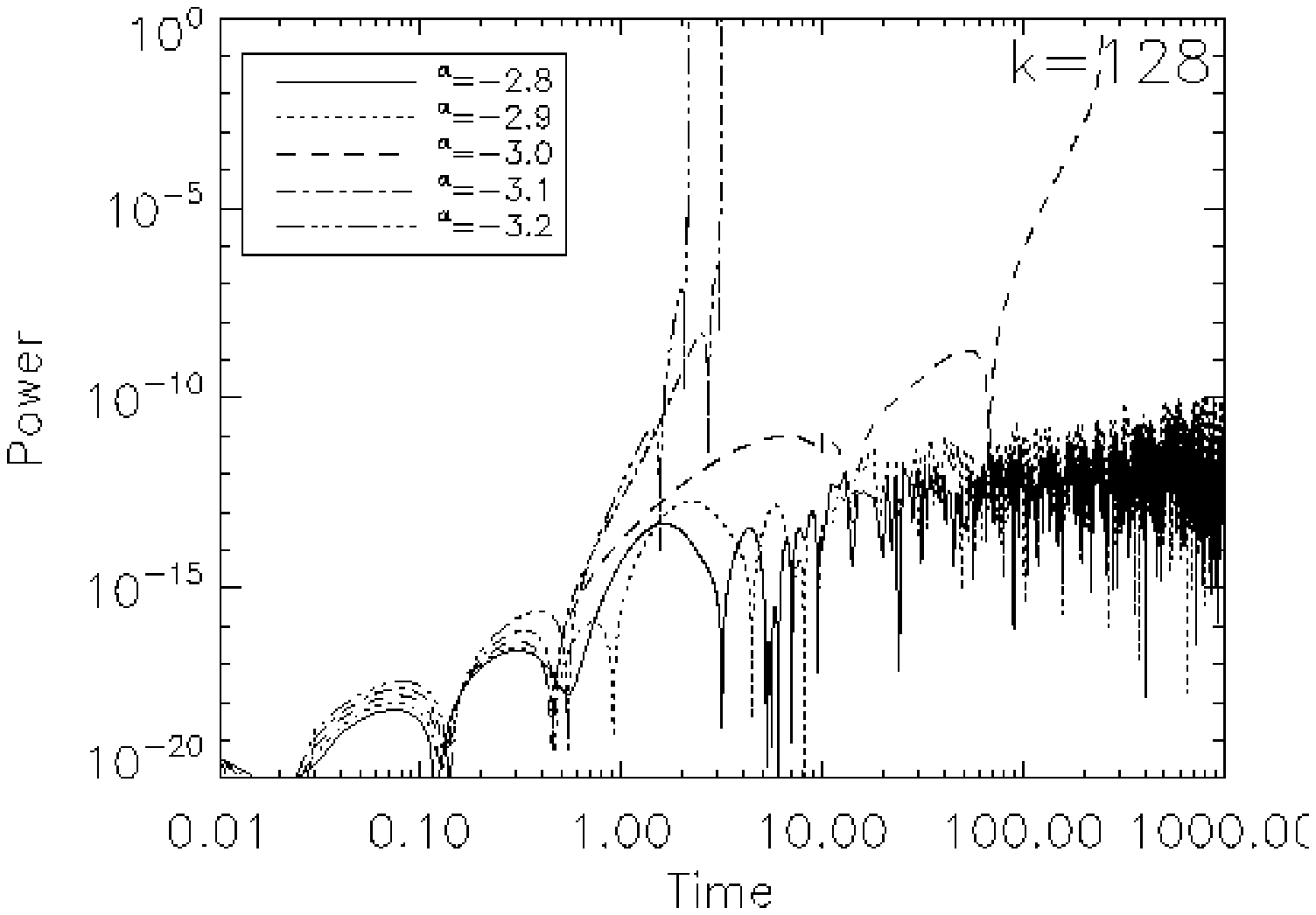}
  \caption{The evolution of the power at modes with $k = 16$, $32$, $64$, and $128$ (and $m = 0$) of the density for the
           study of Rayleigh's criterion discussed in \S\ref{sec:tests_rayleigh}.}\label{fig:rayleigh}
\end{figure*}

We now study the ability of our algorithm to model correctly instabilities in hydrodynamic flows. In the absence of
viscosity, the necessary and sufficient condition for a rotating flow with angular velocity $\Omega(r)$ to be stable is
\begin{equation}
  \frac{d}{dr}(r^2\Omega)^2 > 0, \label{eq:rayleigh}
\end{equation}
which is known as Rayleigh's criterion \citep{Chandrasekhar1981}. Although Rayleigh's criterion usually describes the
stability condition in a Couette flow, i.e., for an incompressible fluid between two rotating cylinders, the criterion
is also valid for rotating compressible fluids, when the unperturbed radial velocity vanishes everywhere in the flow.
This can be achieved by balancing the centrifugal and centripetal accelerations. Let
\begin{equation}
  \kappa^2 = 2\Omega\left(2\Omega + r\frac{d\Omega}{dr}\right)
\end{equation}
be the square of the epicyclic frequency. Rayleigh's criterion is equivalent to requiring $\kappa^2 > 0$, because
$d(r^2\Omega)^2/dr = r^3\kappa^2$.

We define a modified Newtonian gravity with a negative gravitational index $\alpha < 0$ such that the gravitational
acceleration is
\begin{equation}
  g_r(r) \equiv -\frac{GM}{r^{|\alpha|}}.
\end{equation}
Setting $GM = 1$, the corresponding ``Keplerian'' velocity $v_\phi$ is given by
\begin{equation}
  v_\phi = \sqrt{r^{\alpha+1}}. \label{eq:velocity}
\end{equation}
Substituting this into Rayleigh's criterion, i.e., equation~(\ref{eq:rayleigh}), we find that $\alpha = -3$ is the
critical stability condition. The growth rate of the instability is
\begin{equation}
  \tau = \mathrm{Im}\sqrt{\kappa^2} = \mathrm{Im}\sqrt{(\alpha+3)r^{\alpha-1}}.
\end{equation}
Of course, the same result can be obtained by observing that the effective potential
\begin{equation}
  V_\mathrm{eff}(r) = \frac{L^2}{2r^2} + \frac{1}{(\alpha+1)}r^{\alpha+1},
\end{equation}
has no local minimum when $\alpha < -3$; here $L$ is the angular momentum of a fluid element with a unit mass.

We use a spatial resolution of $257\times64$ collocation points to test the Rayleigh criterion. The initial conditions
are with a uniform density, $\Sigma_0 = 1$, and the velocity profile~(\ref{eq:velocity}). Random perturbations that are
of order $10^{-6}$ are added to all physical quantities and the boundary conditions are set to $v_r|_{r_{\min}} =
v_r|_{r_{\max}} = 0$.

Because Rayleigh's criterion refers to an one-dimensional problem, we discuss here explicitly the evolution of the
$m=0$ mode. This is equivalent to averaging over the azimuthal direction. Figure~\ref{fig:rayleigh} shows the evolution
of power at $k = 16$, $32$, $64$, and $128$ (and $m = 0$) of the density, i.e., it is the plot of
$|\check\Sigma_{k,m}(t)|^2$ against time. When the flow is unstable, the perturbations grow very fast and diverge. As
required by Rayleigh's criterion, flows with $\alpha > -3$ are stable to perturbations.

\subsection{Propagation of Wavefronts}

\begin{figure*}
  \plotone{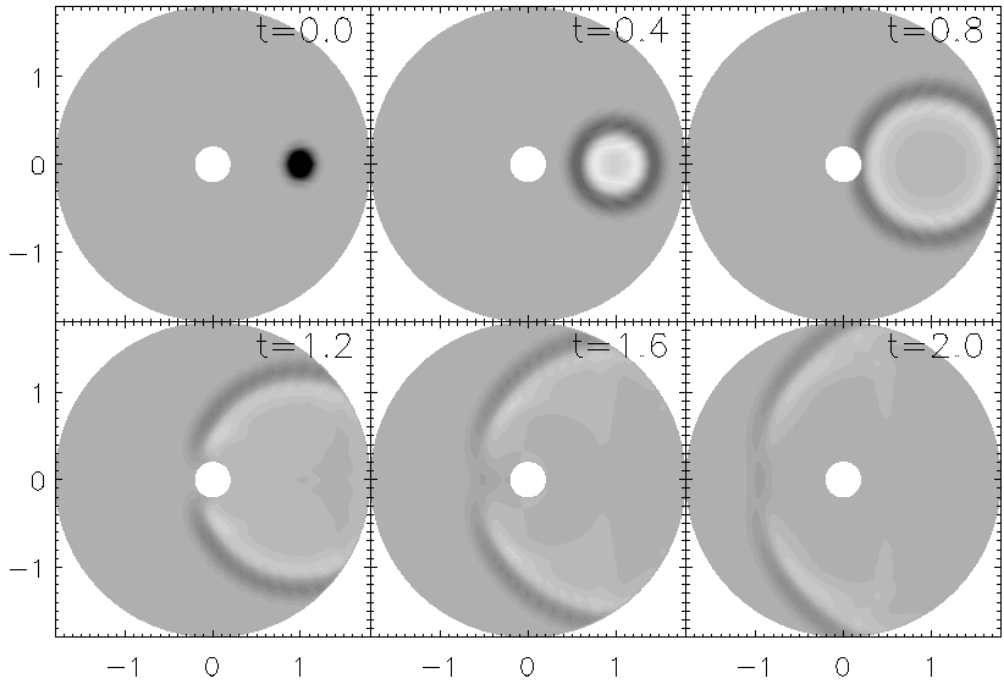}
  \caption{The propagation of a sound wave in a uniform static background. The gray-scale snapshots correspond to the
           fluid density; dark is high and light is low density.}\label{fig:density_contour}
\end{figure*}

\begin{deluxetable}{cccc}
  \tabletypesize{\footnotesize}
  \tablewidth{0pt}
  \tablecaption{Numerical Sound Speeds for the Problem discussed in \S4.4.\label{tab:error_speed}}
  \tablehead{ $\Gamma$ & Numeric Sound Speed & Analytical Sound Speed & Percentage Error }
  \startdata
    1.0  & 1.011210 & 1.000000 & 1.12095\% \\
    1.1  & 1.062727 & 1.048808 & 1.32702\% \\
    1.2  & 1.107763 & 1.095445 & 1.12449\% \\
    1.3  & 1.153139 & 1.140175 & 1.13695\% \\
    1.4  & 1.193408 & 1.183216 & 0.86140\% \\
    1.5  & 1.235294 & 1.224744 & 0.86136\%
  \enddata
\end{deluxetable}

In order to test our implementation of the thermal pressure and the numerical spreading of wavefronts, we neglect
gravity, set all velocities equal to zero, and simulate the propagation of a sound wave. We use the equation of state
$P = K\Sigma^\Gamma$ with $K = 1$ and vary the polytropic index $\Gamma$ from $1$ to $1.5$. This pressure term
decouples the energy equation from the Navier-Stokes equation and the sound speed is simply
$\sqrt{K\Gamma\Sigma^{\Gamma-1}}$. The initial density is
\begin{equation}
  \Sigma_0(r,\phi) = 1 + 10^{-6}\exp\left[-60(1 + r^2 - 2r\cos\phi)\right].
\end{equation}
We use $257\times64$ collocation points and calculate the solution from $t = 0$ to $2$.

Figure~\ref{fig:density_contour} shows six gray-scale snapshots of the density for $\Gamma = 1$. The resulting sound
wave propagates outward as a circle. Note that, because of our boundary treatment, the density wave is not reflected
back as it propagates close to the boundaries. To measure the sound speed, we take the density profile at $\phi = 0$
and trace the wavefronts. The collocation points that correspond to the peak density are then used to calculate, via
$\chi^2$-fitting, the sound speed. Table~\ref{tab:error_speed} shows the $\chi^2$-fitted sound speed of the simulation.
Compared to the analytical sound speed, the error is of order 1\%.

\subsection{Viscous Spreading of a Fluid Ring}\label{sec:tests_pringle}

\begin{figure}
  \plotone{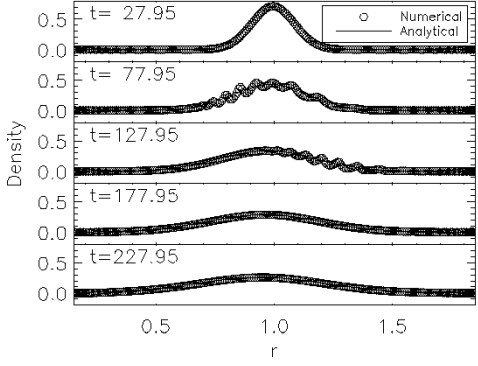}
  \caption{The evolution of a viscous spreading ring. The solid line is the analytical solution and the circles are
           numerical solutions. Note that concentric rings appear at the beginning, but, after about 24 ring rotations,
           they decay away and we recover the analytic solution.}\label{fig:pringle}
\end{figure}

The spreading of an axisymmetric viscous ring with small kinematic viscosity $\nu_\mathrm{s}$
\citep{Pringle1981,Frank2002} has become a standard test for numerical methods including both smoothed particle
hydrodynamics (SPH) and grid-based codes. It has also often been used to guide studies of many other astrophysical
problems, such as the fall-back disk after a supernova explosion \citep[see, e.g.,][]{VanParadijs1995,Eksi2003}.

In the standard solution, it is assumed that $v_\phi = r\Omega_K \gg v_r$, so the viscosity tensor can be approximated
by
\begin{equation}
  \tau_{rr} = \tau_{\phi\phi} = 0,\ \ \
  \tau_{r\phi} = \tau_{\phi r} = \nu_\mathrm{s}\Sigma\left(\frac{\partial v_\phi}{\partial r} - \frac{v_\phi}{r}\right).
\end{equation}
Using conservation of mass, conservation of angular momentum, the fact that $v_\phi$ is always close to the Keplerian
value, and assuming that the kinematic viscosity coefficient $v_\mathrm{s}$ is a constant, the analytical solution for
the density, in our units, becomes
\begin{equation}
  \Sigma(r,t) = \frac{1}{12\pi\nu_\mathrm{s}r^{1/4}t}\exp\left(-\frac{1+r^2}{12\nu_\mathrm{s}t}\right)
  I_{1/4}\left(\frac{2r}{12\nu_\mathrm{s}t}\right), \label{eq:pringle}
\end{equation}
where $I_{1/4}$ is a modified Bessel function.

We perform our simulations with axisymmetric initial conditions. We follow \citet{Speith2003} and use a typical value
$\nu_\mathrm{s} = 4.77\times10^{-5}$. The energy equation is decoupled from the hydrodynamic equations as in the
previous subsection. We set $\Gamma = 1$, and choose $K$ such that $c_\mathrm{s} = 10^{-8}$. The spatial resolution is
$257\times64$ and the time step factor $\delta = 0.8$. We start with an initial density given by
equation~(\ref{eq:pringle}) at $t = 27.95$, which corresponds to $12\nu_\mathrm{s}t = 0.016$. The boundary conditions
are $v_\phi|_{r_{\max}} = (r\Omega_K)|_{r_{\max}}$ and $v_\phi|_{r_{\min}} = (r\Omega_K)|_{r_{\min}}$, which are the
ones assumed in the standard solution.

Figure~\ref{fig:pringle} compares the analytical solution to the numerical solution.  The simulation shows that some
concentric rings appear at the beginning. They come from the relaxation effect as our initial velocities approach
standard solution. In fact, these additional structures have been found in other simulations \citep[for more
information see][]{Monaghan1992}, and decay away after about 24 rotations.

\subsection{Shakura-Sunyaev Steady Disk Solution}

\begin{figure*}
  \plottwo{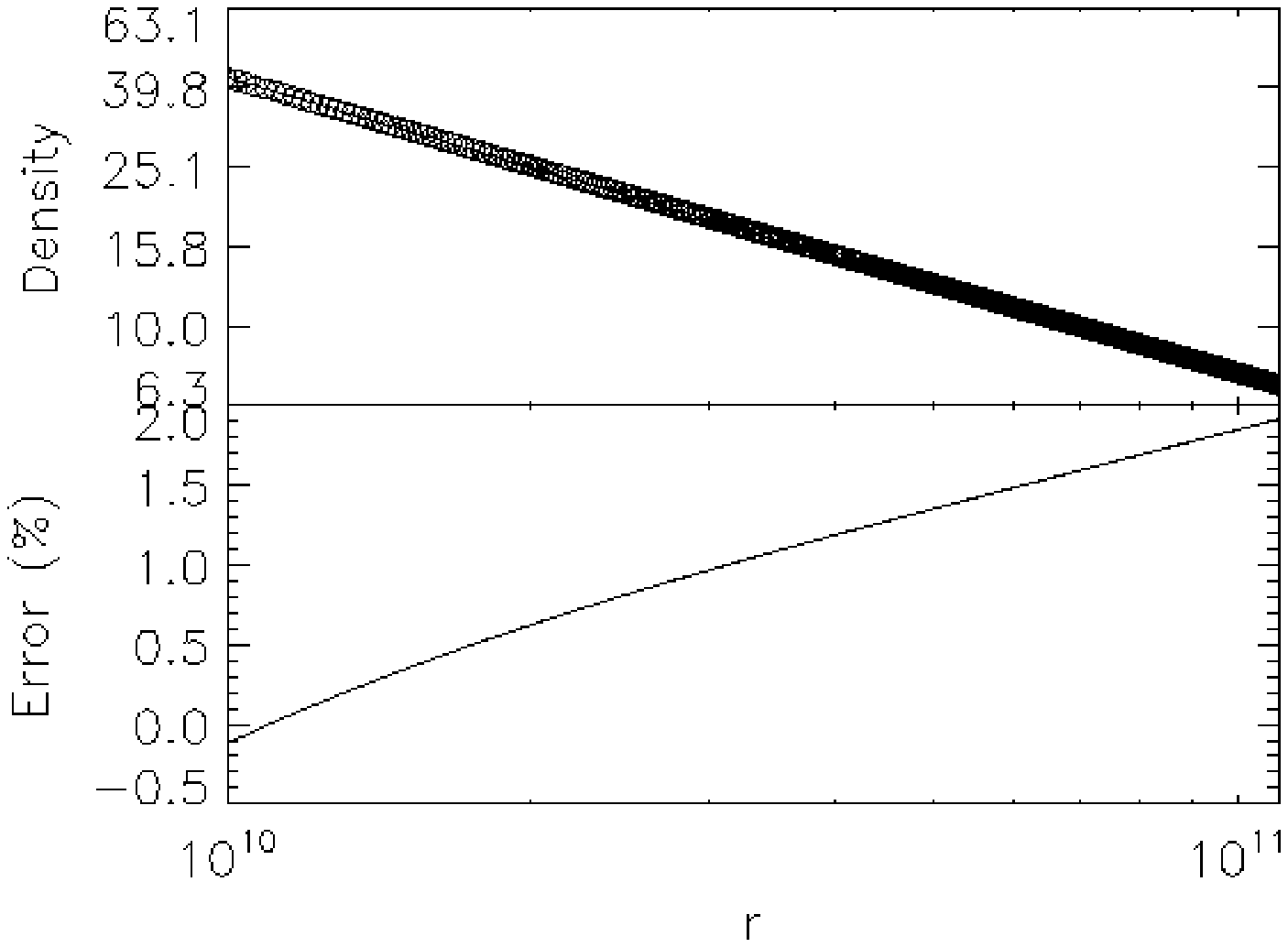}{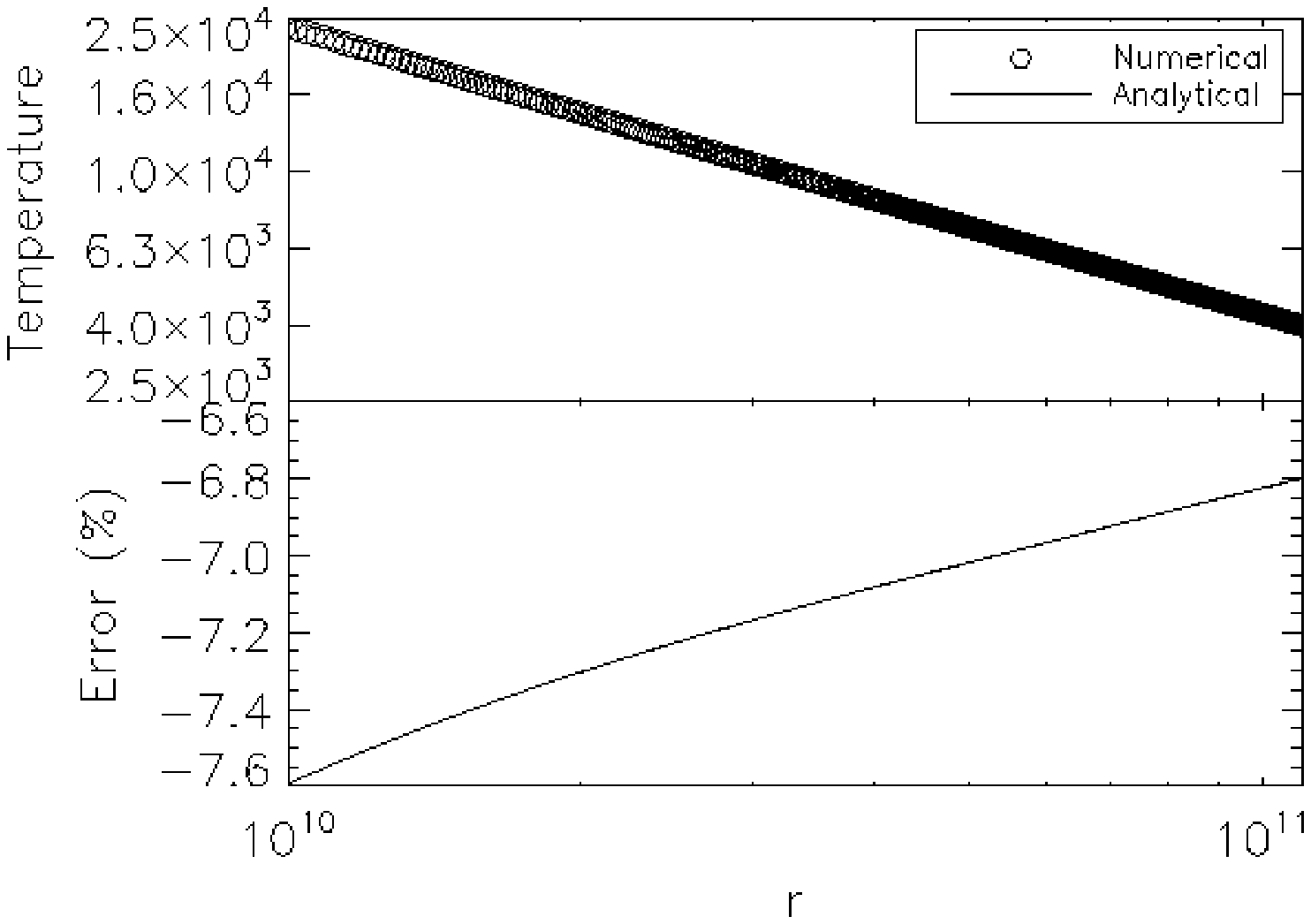} \\
  \plottwo{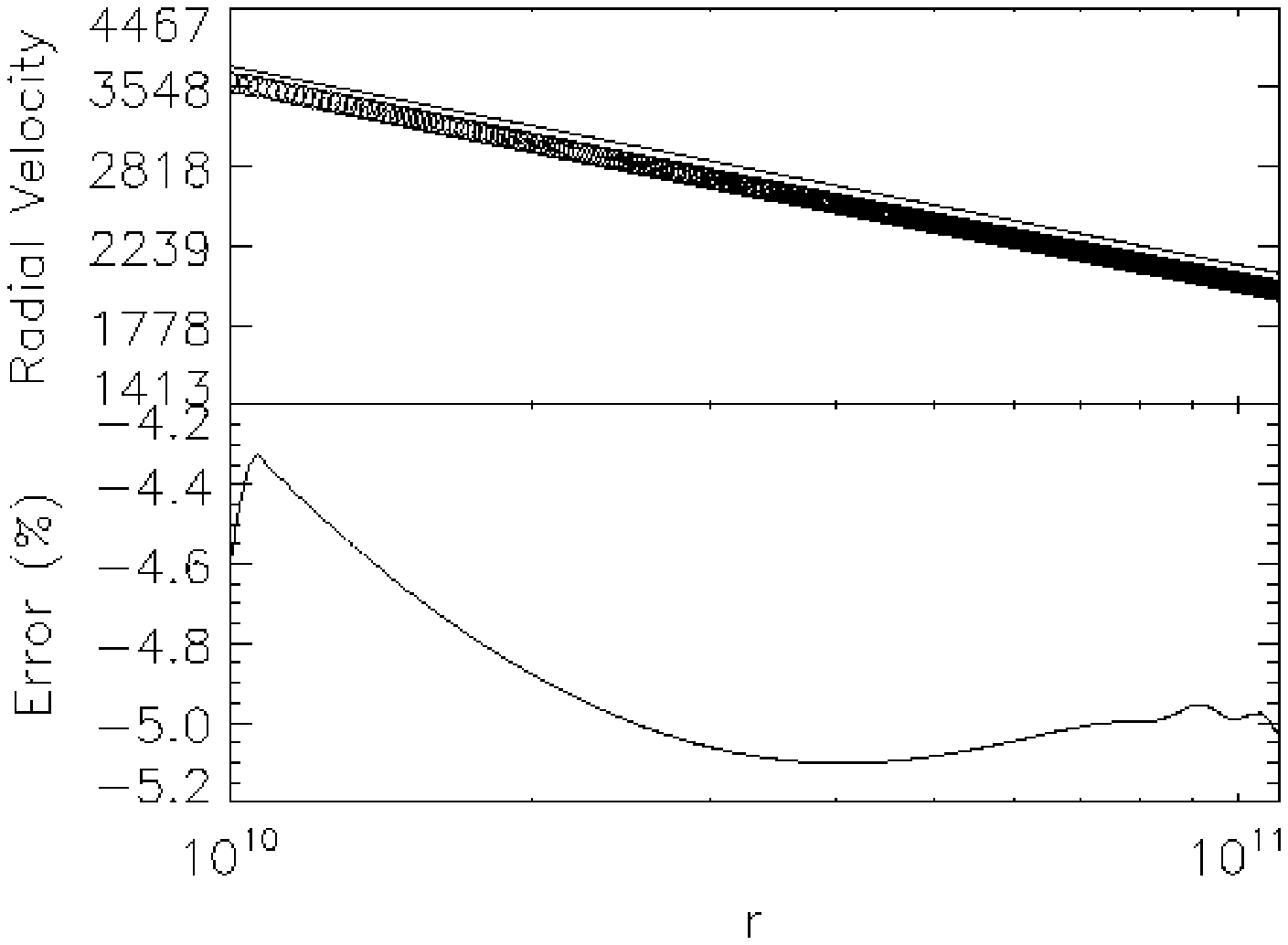}{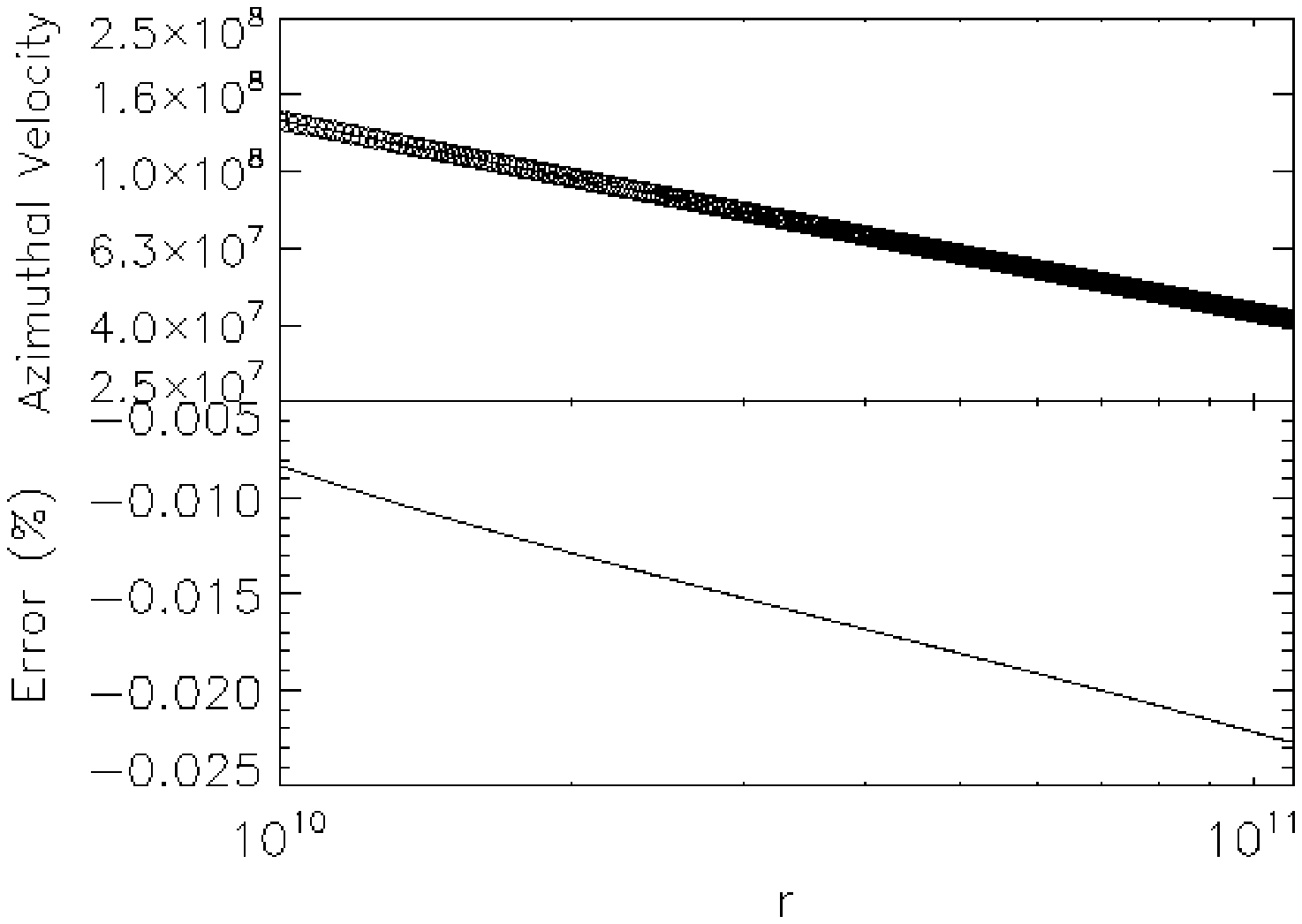}
  \caption{\emph{(Top)} The analytical solution (solid line) compared to the numerical solution (open circles) of the
           Shakura-Sunyaev steady disk solution in log-log scale.
           \emph{(Bottom)} The percentage numerical errors, defined as $(f_\mathrm{num}/f_\mathrm{ana} - 1)\times100\%$.
           All the units along the vertical axis are cgs units, i.e., $\mathrm{g\;cm}^{-2}$ for density, $\mathrm K$ for
           temperature, and $\mathrm{cm\;s}^{-1}$ for velocities.}
           \label{fig:shakura_sunyaev}
\end{figure*}

Here we consider the thin disk problem and suppose that the disk is able to settle to a steady-state structure. By
assuming an $\alpha$-viscosity law, \citet{Shakura1973} solved the local disk structure analytically. Their work has
been highly cited for studies in accretion disks. (See Frank, King, \& Raine 2002 and reference therein).

In our simulation, we solve the hydrodynamic equations~(\ref{eq:continuity}) -- (\ref{eq:energy}) including the energy
equation. The kinematic viscosity coefficient is given by
\begin{equation}
  \nu_\mathrm{s} = \alpha c_\mathrm{s} H.
\end{equation}
The sound speed is given by $c_\mathrm{s}^2 = \partial P/\partial \Sigma = k_\mathrm{B}T/\mu m_\mathrm{H}$ and the disk
scale height is $H = r c_\mathrm{s}/v_\phi$. We, also, assume the vertical optical depth of the disk to be
$\tau_\mathrm{d} = \Sigma\kappa_\mathrm{R}$, where the Rosseland mean opacity $\kappa_\mathrm{R}$ is given by Kramers'
law\footnote{See again \citet{Frank2002}, p.94 for notes on errors in the literature.}
\begin{equation}
  \kappa_\mathrm{R} = 6.07\times10^{22}\frac{\Sigma}{\sqrt{4\pi}H} T^{-7/2}\mathrm{cm^2\;g^{-1}},
\end{equation}

Based on the above assumption and using the fact that $\nu_\mathrm{s}$ is small, the Shakura-Sunyaev disk solution is
given by
\begin{eqnarray}
  \Sigma & = &  6.15\;        \alpha^{-4/5}\dot M_{16}^{7/10}m_1^{ 1/4}R_{10}^{-3/4}f^{14/5} \mathrm{g\;cm^{-2}}, \\
  v_r    & = & -2.59\times10^4\alpha^{ 4/5}\dot M_{16}^{3/10}m_1^{-1/4}R_{10}^{-1/4}f^{-14/5}\mathrm{cm\;s^{-1}}, \\
  v_\phi & = &  1.15\times10^8                               m_1^{ 1/2}R_{10}^{-1/2}         \mathrm{cm\;s^{-1}}, \\
  T      & = &  1.48\times10^4\alpha^{-1/5}\dot M_{16}^{3/10}m_1^{ 1/4}R_{10}^{-3/4}f^{6/5}  \mathrm{K},
\end{eqnarray}
where $\dot M_{16} = \dot M/10^{16}\mathrm{g\;s^{-1}}$, $m_1 = M/M_\sun$, $R_{10} = r/(10^{10}\mathrm{cm})$, and $f =
(1 - \sqrt{R_*/r})^{1/4}$, with $R_*$ being the radius of the central object. Considering the flow around a standard
neutron star with a mass of $1.4M_\sun$, and a radius of $8.5\times10^5\mathrm{cm}$, we setup the simulation with
$r_{\min} = 10^{9}\mathrm{cm}$ and $r_{\max} = 1.2\times10^{11}\mathrm{cm}$. The initial conditions for density and
energy are as the Shakura-Sunyaev disk solution with $\dot M_{16} = 1$, $m_1 = 1.4$, and $\alpha=0.1$. We set the
initial radial velocity to zero and the initial azimuthal velocity to be Keplerian. Perturbations were added to all
variables with amplitude $0.1$ and order $3$. The simulation runs with $513\times64$ collocation points up to $t =
10^7\mathrm{s}$, which is approximately $16$ viscous time scales.

In Figure~\ref{fig:shakura_sunyaev}, we show our numerical steady solution for $\alpha = 0.1$, at $t = 10^7\mathrm{s}$;
note that we use the magnitude of the radial velocity in order to plot it in log-log scale. The numerical values are
different from the Shakura-Sunyaev solution by a few percent. We point out explicitly that the pressure term in the
Navier-Stokes equation is neglected in obtaining the analytical solution. Although the pressure term does not transport
angular momentum, the effect of pressure compared to those of viscosity is proportional to $1/\alpha$. Since $\alpha =
0.1$ in our simulation, the pressure effects can account for the difference between the analytical and numerical
solution.

\section{Application: Non-Axisymmetric Instabilities in Viscous Spreading Rings}

\begin{figure*}
  \plottwo{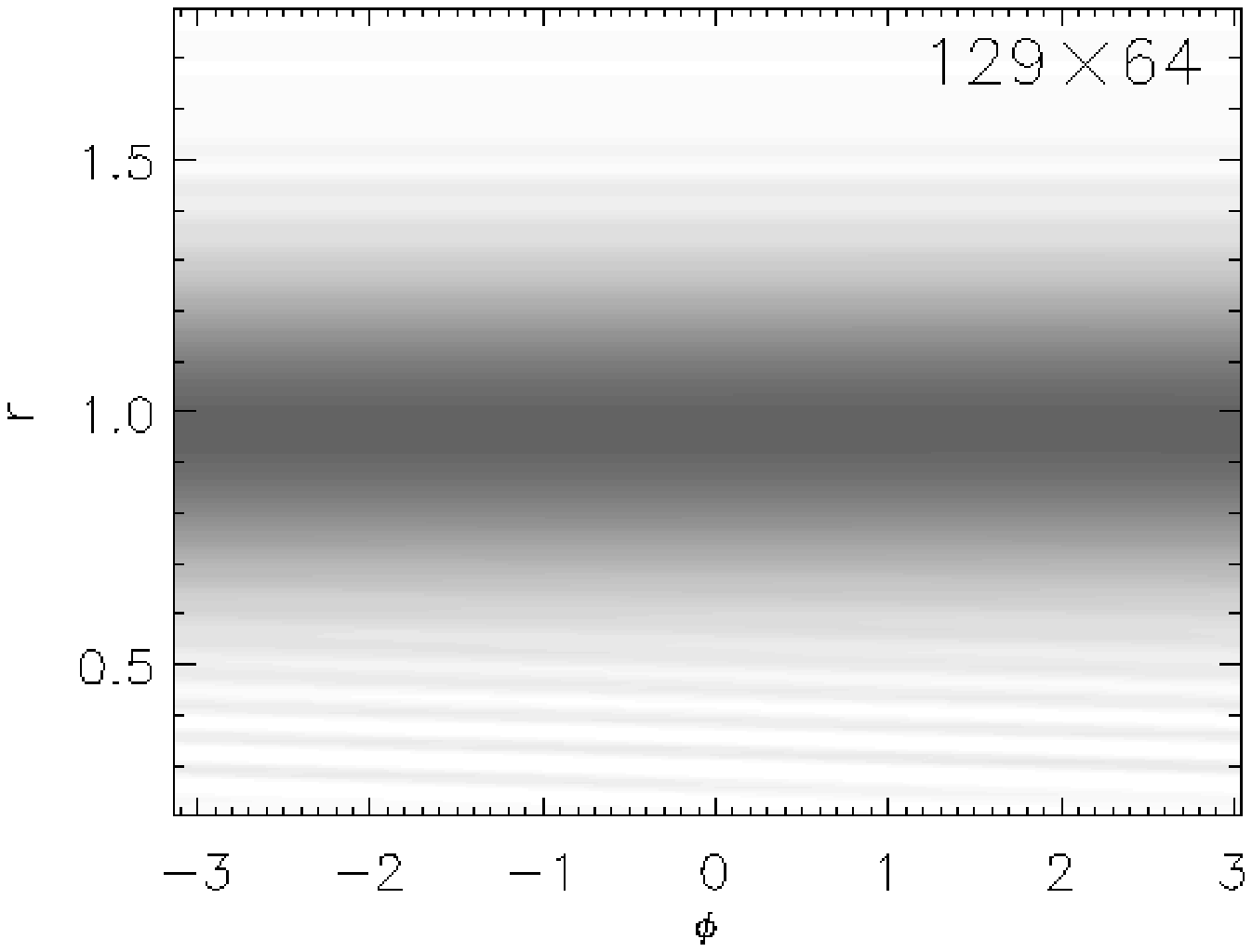}{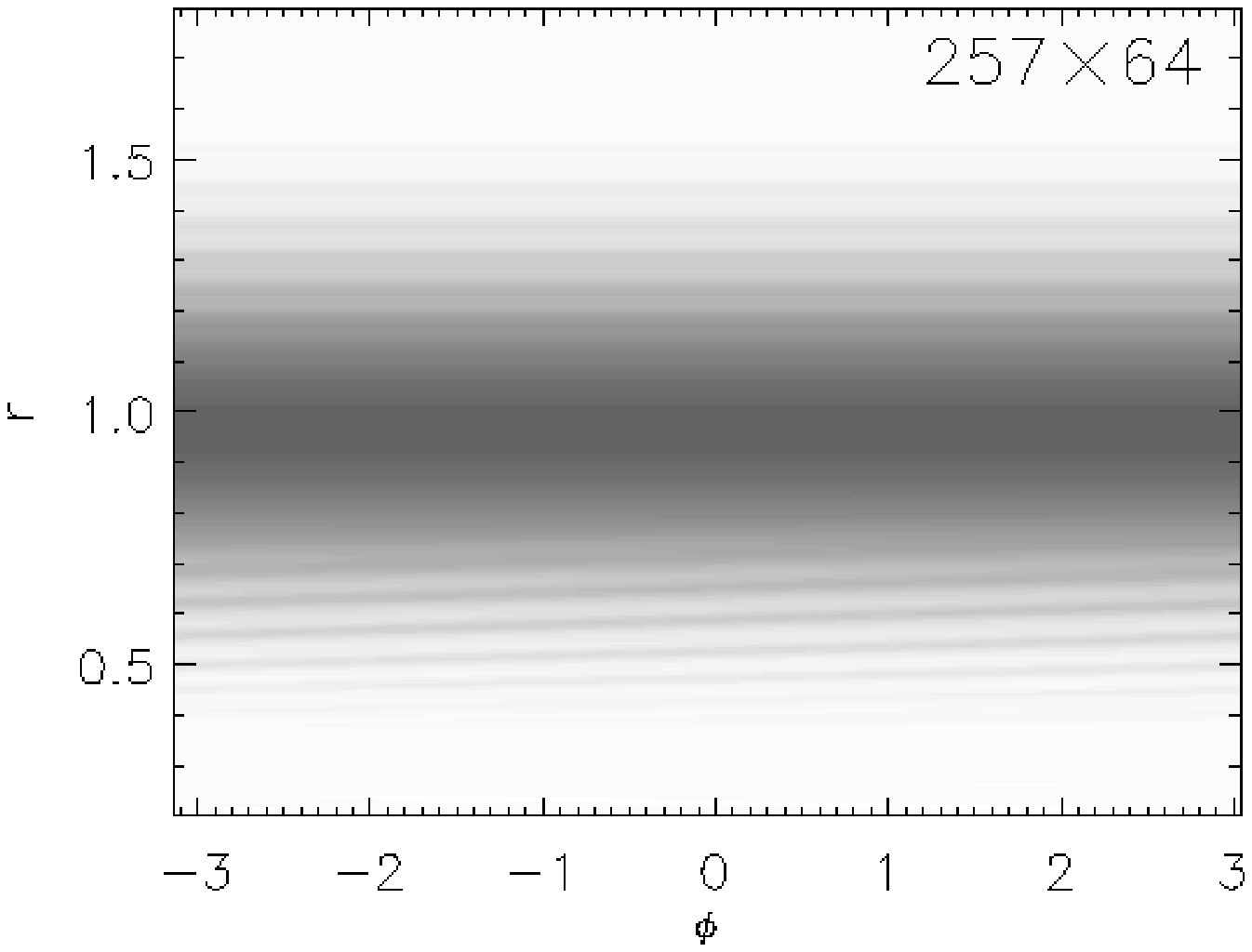} \\
  \plottwo{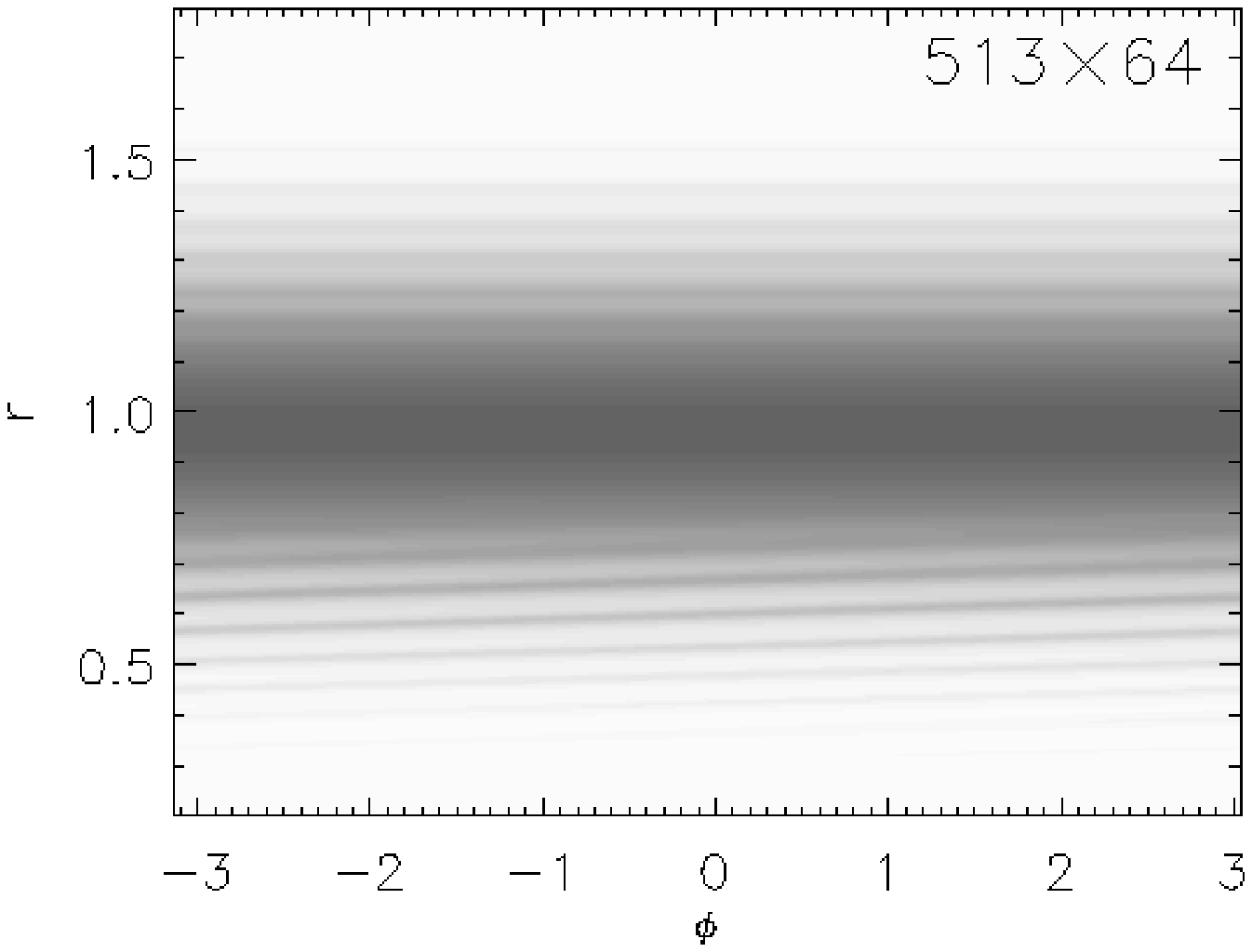}{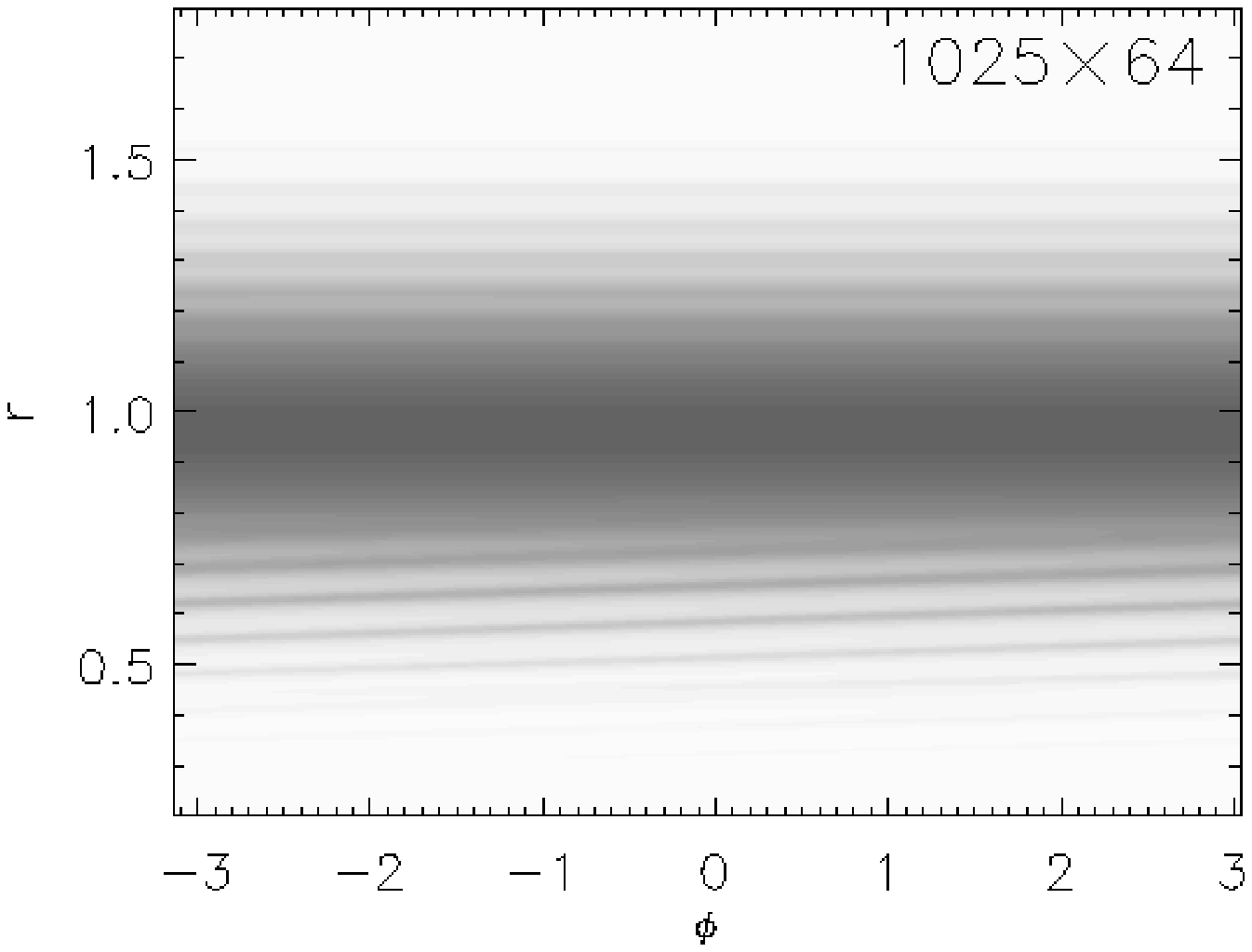}
  \caption{Snapshot of simulations of a viscous spreading ring with different resolutions. Note that for different
           resolution, the direction of the spirals are different.}\label{fig:speith_kley_compare}
\end{figure*}

\begin{figure*}
  \plotone{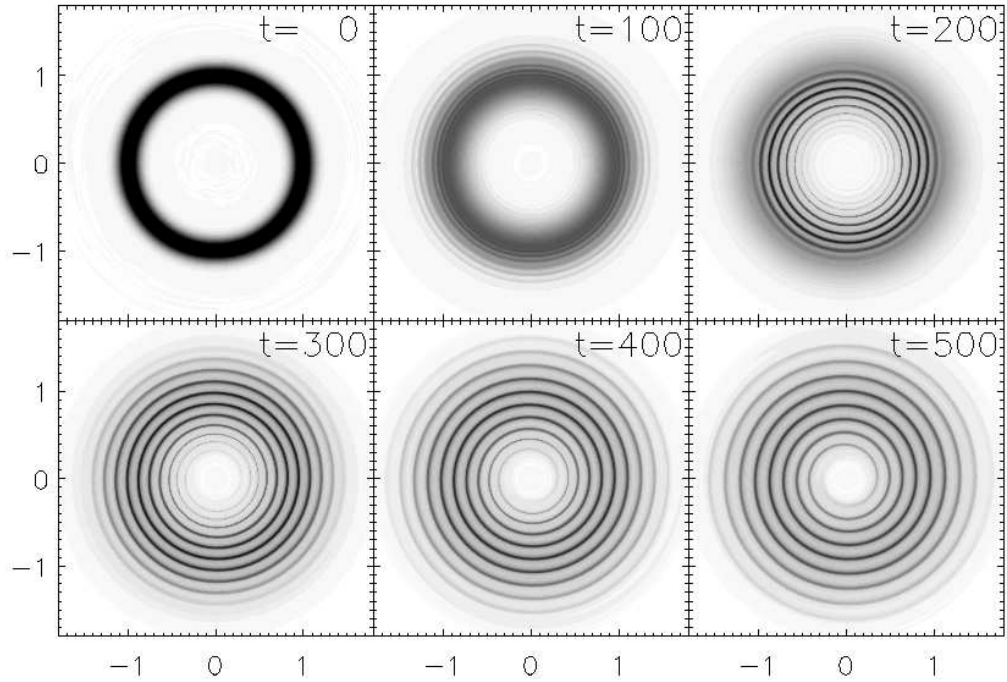}
  \caption{The evolution of a viscous spreading ring with non-axisymmetric perturbations. At $t = 100$, there are
           concentric rings due to the initial relaxation. At $t = 200$ it is clear that a small, one-armed spiral
           starts to develop in the inner region. The remaining figures, then, show the development of the one-armed
           spiral.}\label{fig:speith_kley}
\end{figure*}

\begin{figure}
  \plotone{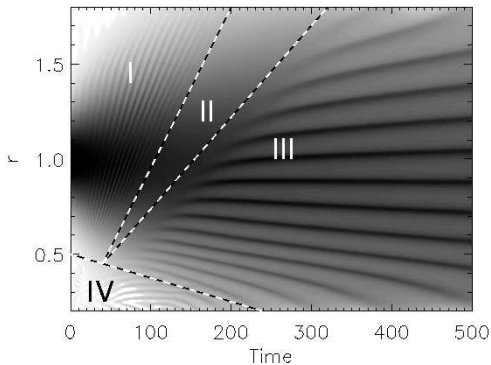}
  \caption{The contour plot of density at $\phi = 0$ against time. Region I shows the relaxation effect. Region II shows
           that the solution has relaxed to the standard solution. The Speith-Kley instability develops in region III.
           In region IV, the unstable wavelengths are too short to be resolved in our numerical method.}
           \label{fig:speith_kley_evolve}
\end{figure}

\begin{figure}
  \plotone{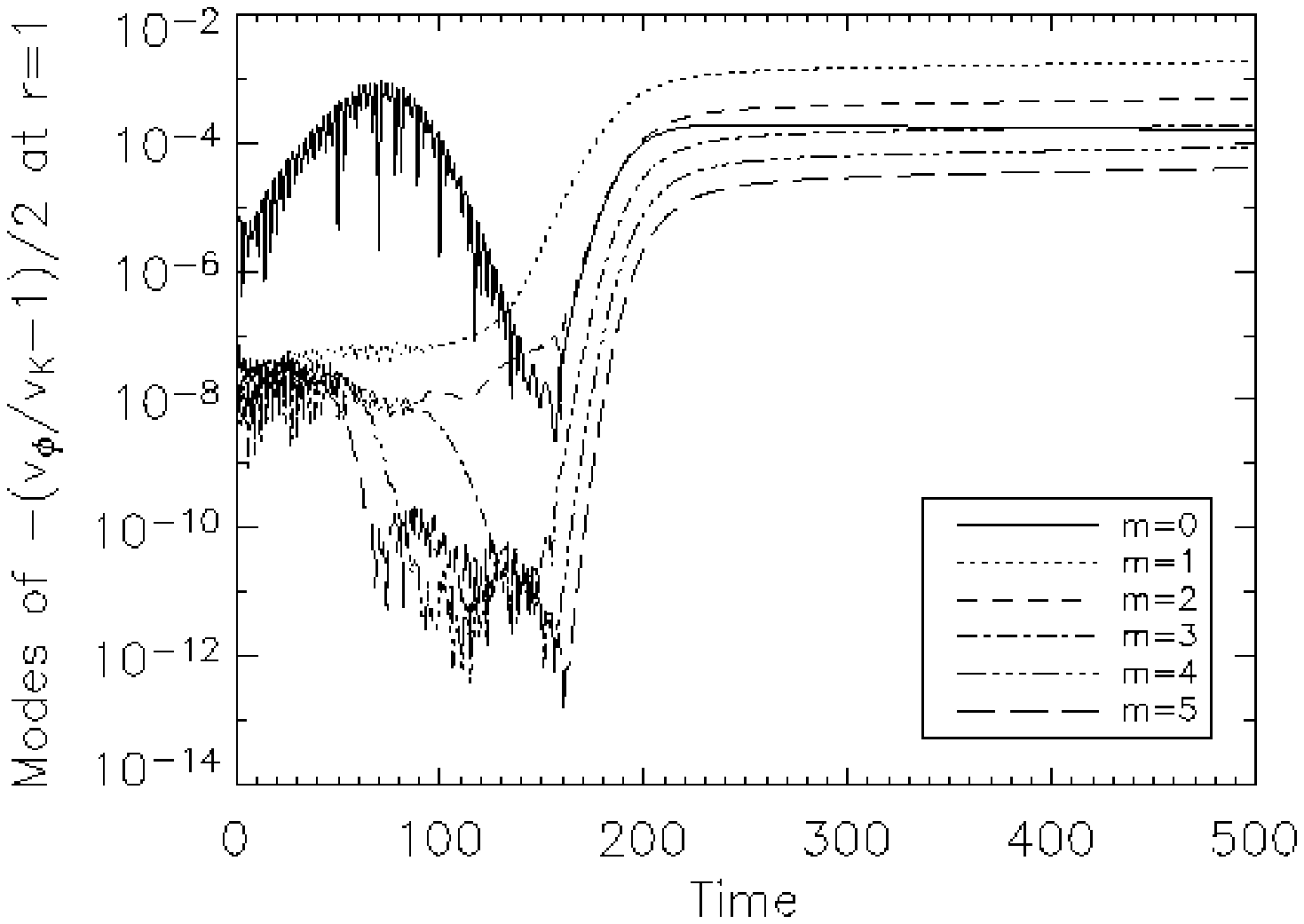}
  \caption{The evolution of the first 6 modes of $\hat E(r = 1,t)$.}
           \label{fig:speith_kley_modes_evolve}
\end{figure}

As we mentioned in \S\ref{sec:tests_pringle}, the viscous spreading ring has become a standard test for numerical
simulations. In these simulations, additional structures always appear, such as the concentric rings in
Figure~\ref{fig:pringle}. We attributed the presence of these concentric rings to relaxation effects of the initial
conditions. \citet{Maddison1996}, on the other hand, showed that the concentric rings that appear in SPH simulations
using the artificial viscosity given by \citet{Monaghan1992} are numerical artifacts. Later, \citet{Speith2003} used
perturbation methods to show that viscous spreading rings are unstable to non-axisymmetric perturbations.

Assuming a compressible, pressureless, and viscous fluid, the equations used in \citet{Speith2003} are the continuity
equation~(\ref{eq:continuity}) and the Navier-Stokes equation~(\ref{eq:navier_stokes}), with zero pressure. By setting
$\nu_\mathrm{s} = \mu_\mathrm{s}/\Sigma$ to be constant and $\mu_\mathrm{r} = \mu_\mathrm{b}= 0$, the first order
instability is an one-armed spiral, whereas the second order instability is a superposition of an one-armed and a
two-armed spirals.

If we define a function
\begin{equation}
  \hat E(t,r) = 2- \frac{2v_\phi}{r\Omega_\mathrm{K}},
\end{equation}
where $\Omega_\mathrm{K} = \sqrt{GM/r^3}$ is the Keplerian angular velocity, then $\hat E(t,r)$ measures the first
order perturbation amplitude. Its dispersion relation is
\begin{equation}
  \sigma_\mathrm{d} = -k\left(\frac{\nu_\mathrm{s}}{r}
  + \frac{8}{3}\frac{\nu_\mathrm{s}}{\hat\Sigma_0}\frac{\partial\hat\Sigma_0}{\partial r}\right)
  +i\left(k^2\frac{\nu_\mathrm{s}}{3}
  - \frac{\nu_\mathrm{s}}{r\hat\Sigma_0}\frac{\partial\hat\Sigma_0}{\partial r}\right)
\end{equation}
where $k$ is the radial wavenumber and $\hat\Sigma_0$ is the standard solution~(\ref{eq:pringle}). Assuming
$12\nu_\mathrm{s}t \ll r$, it follows that the condition for instability is
\begin{equation}
  k^2 > \frac{3}{r\hat\Sigma_0}\frac{\partial\hat\Sigma_0}{\partial r}
\end{equation}
and the growth rate is
\begin{equation}
  \mathrm{Im}(\sigma_\mathrm{d}) = \frac{1}{3}k^2\nu_\mathrm{s} + \frac{3\nu_\mathrm{s}}{4r^2} + \frac{1}{6t}(1 - r).
  \label{eq:growth_rate}
\end{equation}
In order to test this claim, we carried out a test with the same condition as Speith and Kley's RH2D simulation
\citep[a radiative hydrodynamic code; see][]{Kley1989}. The only differences are in the initial perturbations and a
small constant background term in the density, i.e., the setup is the same as in \S\ref{sec:tests_pringle} but we add
initial perturbations by the method described in \S\ref{sec:perturbation} with order 3 and magnitude 0.001. The
additional constant background term in the density varies from 0.1 to 0.001 depending upon the resolution, which keeps
the solution stable. We perform the simulation with resolution $129\times64$, $257\times64$, $513\times64$, and
$1025\times64$.

Figure~\ref{fig:speith_kley_compare} shows the density contours for different resolutions at time $t = 120$. It is
clear that the direction of the spiral in low resolution is different then in high resolution. We confirm this result
with Speith and Kley (private communication), i.e., that the transition radius between leading and trailing spirals
vary as the resolution changes. Figure~\ref{fig:speith_kley} is the density contour of the simulation with
$513\times64$ collocation points and 0.001 background density, at different times. At $t = 100$, some concentric rings
due to relaxation effects are seen. At $t = 200$, a one-armed spiral starts to form at the inner disk. The spiral is
fully developed by $t = 300$.

In order to find out the unstable wave number, we plot in Figure~\ref{fig:speith_kley_evolve} the contour of density
along $\phi = 0$ against time. Region I shows the relaxation effect, where the relaxation waves propagate outwards. In
region II, the solution has relaxed to the standard solution. Later on, the Speith-Kley instability starts to develop,
which is shown in region III. We can estimate the wavelength of the fully developed spirals, which varies from $0.1$ to
$0.15$. For region IV, the unstable wavelengths are too short to be resolved in our numerical method.

Finally, in Figure~\ref{fig:speith_kley_modes_evolve} we plot the first 6 modes, i.e., $m = 0, 1, \dots, 5$, of $\hat
E(r = 1, t)$. The late time behavior of our solution agrees with Figure~8 in \citet{Speith2003}. To compare our
numerical result with the analytical growth rate~(\ref{eq:growth_rate}), we $\chi^2$-fit the slope of the $m=1$ mode in
Figure~\ref{fig:speith_kley_modes_evolve} between $t = 250$ and $t = 500$, which gives $\mathrm{Im}(\sigma_\mathrm{d})
\approx 0.0013$. That corresponds to unstable wavenumber $k \approx 9.0$. It agrees with the width of the spiral
$\approx 0.11$ as shown in Figures~\ref{fig:speith_kley} and \ref{fig:speith_kley_evolve}.

Physically, the viscous spreading ring problem is an important guide to understanding the transport of angular momentum
in accretion disks. We have independently verified the Speith-Kley instability: a cold viscous disk is unstable to
non-axisymmetric perturbations.

%---------------------------------------------------------------------------------------------------------------------

\section{Conclusions}

We described a numerical method based on a pseudo-spectral algorithm for studying the rapid variability properties of
two-dimensional, viscous, hydrodynamic accretion disks. We demonstrated the ability of the spectral methods to handle
correctly non-reflective boundary conditions and different stability conditions. We verified the implementation of the
algorithm using various test problems. Also, we confirmed the non-axisymmetric instability of viscous spreading rings
discovered by Speith and Kley.

Spectral methods calculate the time evolution of  a solution as an interaction of different waves. From a mathematical
point of view, this approach naturally agrees with existing methods to study non-linear equations. It helps us to
confirm directly the result of linear mode analysis as well as turbulence theories. From a computational point of view,
these methods are high order and produce accurate result in most problems. It is straightforward to extend the current
algorithm to simulate 3D MHD accretion disks. In particular, because of the high order of the spectral methods, we can
solve directly for the vector potential and not for the magnetic field, in order to guarantee that the resulting field
will be free of divergence at each time step. Also, spectral methods can be used to solve self-gravity problem without
increasing significantly the computation time. Finally, and most important, form a physical point of view, our
algorithm provides another tool to set up numerical experiments and test astrophysical models. It is useful to confirm
independently the results from other numerical simulations as well as test new models.

%---------------------------------------------------------------------------------------------------------------------

\begin{appendix}

\section{Convection-Conservative Mixed Formulism}\label{app:equations}

As we describe in \S\ref{sec:non-linear}, we employ a convective-conservative mixed method. The non-linear term
$(\mathbf{v}\cdot\nabla)\mathbf{v}$ in the Euler equation is implemented in convective from, and all other terms are in
conservation form.
\begin{eqnarray}
  \partial_t\Sigma & = & - \frac{\partial_r(r\Sigma v_r)}{r} - \frac{\partial_\phi(\Sigma v_\phi)}{r},
  \label{eq:hydro_begin}\\
  \partial_t v_r & = & - v_r\partial_r v_r - \frac{v_\phi}{r}(\partial_\phi v_r - v_\phi)
  + \frac{\partial_r(r\tau_{rr} - rP)}{r\Sigma} + \frac{\partial_\phi\tau_{r\phi}}{r\Sigma}
  + \frac{P - \tau_{\phi\phi}}{r\Sigma} - g_r, \\
  \partial_t v_\phi & = & - v_r\partial_r v_\phi - \frac{v_\phi}{r}(\partial_\phi v_\phi + v_r)
  + \frac{\partial_r(r\tau_{r\phi})}{r\Sigma} + \frac{\partial_\phi(\tau_{\phi\phi}-P)}{r\Sigma}
  + \frac{\tau_{r\phi}}{r\Sigma}, \\
  \partial_t E & = & - \frac{\partial_r(rE v_r + rq_r + rF_r)}{r} - \frac{\partial_\phi(E v_\phi + q_\phi + F_\phi)}{r}
  - P\left(\partial_r v_r + \frac{\partial_\phi v_\phi}{r} + \frac{v_r}{r}\right) + \phi - 2F_z. \ \ \ \ \ \
  \label{eq:hydro_end}
\end{eqnarray}
Here, the notation $\partial_t$ denote partial derivative with respect to time. The other notations $\partial_r$ and
$\partial_\phi$ denote the numerical derivative operators. Hence, the notation $v_r\partial_r v_r$ means, we first take
a numerical derivative of the variable $v_r$ and multiply the result with $v_r$, which is in convective form. On the
other hand, $\partial_r(r\Sigma v_r)/r$ means, we first use a temporary variable to store the produce $r\Sigma v_r$,
apply the procedure to calculate the numerical derivative on it, and then divide it by $r$. This is the conservative
form.

The viscosity tensor $\tau_{i\!j}$ in the above equation has the following general form
\begin{equation}
  \tau_{i\!j} = 2(\mu_\mathrm{r} + \mu_\mathrm{s})e_{i\!j} + \left(\mu_\mathrm{r} + \mu_\mathrm{b}
  - \frac{2}{3}\mu_\mathrm{s}\right)\nabla\cdot\mathbf{v}.
\end{equation}
As we describe before, $\mu_\mathrm{r}$, $\mu_\mathrm{b}$, and $\mu_\mathrm{s}$ are coefficient of radiative, bulk, and
shearing viscosity. The strain rate tensor $e_{i\!j}$ written in cylindrical coordinate becomes
\begin{eqnarray}
  e_{rr} & = & \partial_r v_r \\
  e_{r\phi} = e_{\phi r} & = & \frac{1}{2}\left(\partial_r v_\phi - \frac{v_\phi}{r}
  + \frac{\partial_\phi v_r}{r}\right) \\
  e_{\phi\phi} & = & \frac{\partial_\phi v_\phi}{r} + \frac{v_r}{r}
\end{eqnarray}
and $\nabla\cdot v$ in cylindrical coordinate is
\begin{equation}
  \nabla\cdot v = \partial_r v_r + \frac{v_r}{r} + \frac{\partial_\phi v_\phi}{r}.
\end{equation}
For completeness, we provide again the viscous dissipation rate
\begin{equation}
  \phi = 2(\mu_\mathrm{r} + \mu_\mathrm{s})(e_{i\!j})^2 + \left(\mu_\mathrm{r} + \mu_\mathrm{b}
  - \frac{2}{3}\mu_\mathrm{s}\right)(\nabla\cdot\mathbf{v})^2.
\end{equation}
The functional forms of $P$, $\mu_\mathrm{r}$, $\mu_\mathrm{b}$, $\mu_\mathrm{s}$, $\mathbf{q}$, $\mathbf{F}$, and
$F_z$ can be change easily depend upon the physical assumption (where the defaults are described in
\S\ref{sec:equations}).

\end{appendix}

%---------------------------------------------------------------------------------------------------------------------

\acknowledgements

C.-K. C. and D. P. acknowledge the support from a NASA ATP grant NAG-513374. F. \"O. acknowledges support by NASA
through Hubble Fellowship grant HF-01156 from the Space Telescope Science Institute, which is operated by the
Association of Universities for Research in Astronomy, Inc., under NASA contract NAS 5-26555.

%---------------------------------------------------------------------------------------------------------------------

\end{document}